\documentclass[11pt,reqno,a4paper]{amsart}
\newtheorem{thm}{Theorem}[section]
\newtheorem{defi}{Definition}[section]

\newtheorem{pr}{Proposition}[section]
\theoremstyle{definition}
\newtheorem{rem}{Remark}[section]

\newcommand{\be}{\begin{equation}}
\newcommand{\ee}{\end{equation}}
\newcommand{\bea}{\begin{eqnarray}}
\newcommand{\eea}{\end{eqnarray}}
\newcommand{\beb}{\begin{eqnarray*}}
\newcommand{\eeb}{\end{eqnarray*}}
\usepackage{amssymb,amsmath, amsfonts,amsthm,setspace,indentfirst,mathrsfs}
\usepackage{xcolor,url}
\usepackage{minibox}

\usepackage{enumitem}
\usepackage{cite}
\numberwithin{equation}{section}
\begin{document}
%
\title[An exploration of the curvature and inheritance properties of the IBH spacetime]{\bf{ An exploration of the curvature and inheritance properties of the Interior black hole spacetime }}

\author[A. A. Shaikh and Kamiruzzaman]{Absos Ali Shaikh$^{* 1}$ and Kamiruzzaman$^2$}
\address{\noindent\newline$^{1,2}$ Department of Mathematics,
	\newline The University of Burdwan, 
	\newline Golapbag, Burdwan-713104,
	\newline West Bengal, India} 
\email{aashaikh@math.buruniv.ac.in$^1$, aask2003@yahoo.co.in$^1$}
\email{kamiruzzaman8145@gmail.com$^2$}

\date{\today}

\dedicatory{}

\begin{abstract}	
In continuation of the study in	\cite{SDHK_interior_2020},  the present article explores the geometric and curvature properties of the interior black hole (briefly, IBH) spacetime. It is shown that in an IBH spacetime the operator $R\cdot C$ and $C\cdot R$ does not commute with each other and infact the commutator $C\cdot R-R\cdot C$ is linearly dependent with $Q(g,R)$ and $Q(S,R)$ as well as $Q(g,C)$ and $Q(S,C)$. Also in IBH spacetime $R \cdot R$ is linearly dependent with $Q(S,R)$ and $Q(g,C)$. It is exhibited that IBH spacetime is $2$-quasi Einstein, Ein$(2)$ and generalized quasi Einstein spacetime in the sense of Chaki, and its conformal $2$-forms are recurrent. We have derived the universal form of the compatible tensors in such a spacetime. We have also demonstrated that the nature of energy momentum tensor of IBH spacetime is pseudosymmetric (see, Theorem $4.1$). Again it is exposed that with respect to the non-Killing vector field $\frac{\partial}{\partial t},$ the IBH spacetime obeys the generalized curvature inheritance, generalized Ricci inheritance, special type of generalized conformal, concircular, conharmonic and generalized Weyl projective inheritance. Finally a comparison between IBH spacetime and KIselev Black Hole (KBH) is displayed.

\end{abstract}

%

\noindent\footnotetext{ $^*$ Corresponding author(Absos Ali Shaikh, E-mail address: aashaikh@math.buruniv.ac.in, aask2003@yahoo.co.in).\\
$\mathbf{2020}$\hspace{5pt}Mathematics\; Subject\; Classification: 53B20; 53B30; 53B50; 53C15; 53C25; 53C35; 83C15. }
\maketitle
%

\section{\bf Introduction}\label{intro}

A black hole is a domain of spacetime where the gravitational force is so intense that neither matter nor electromagnetic energy, such as light can escape. According to Albert Einstein's theory of general relativity a mass that is sufficiently dense can warp spacetime to a black hole, which are objects that have collapsed under their own gravity are characterized by infinite redshift, signifying that they are completely isolated from the rest of the universe. Any event that happen within a black hole cannot affect those taking place outside of it in any manner. Therefore, a significant challenge is to clearly define the boundary separating the inside of the black hole from its outside region.
From the moment they were found as physical phenomenons, black holes have encouraged researchers to explore their fundamental principles of gravity more thoroughly. Though it is difficult to investigate their characteristics, the theoretical knowledge about black hole has raised deep inquiries and it forecasts the essence of space, time and the universe.

 We comprehend the properties of black holes as forecasted by general relativity (GR). A universe with three spatial dimensions and one time dimension is assumed in the majority of black hole research, which makes sense given the state of observations today. Nonetheless, numerous unified theories suggest the existence of a universe with additional dimensions beyond the four we can observe, which makes it beneficial to document higher-dimensional GR black holes for the sake of comparison. Initially, it could be assumed that black holes in higher dimensions are a natural extension of those in 3+1 dimensions, prompting a short examination of black holes in four dimensions.
Uniqueness is a key characteristic for documenting these black holes. In \cite{bb1}, it has been demonstrated that 3+1 black holes are entirely specified by their mass, angular momentum, and electromagnetic charge under mild assumptions like asymptotic flatness and stationarity. As the mass, angular momentum and electric charge are taken into account, the Kerr-Newman (KN) solution \cite{bb2} to the $3+1$ Einstein-Maxwell equations is the most widespread representation of a black hole. Reissner-Nordström \cite{Reis1916,bb5} (mass and charge), Kerr \cite{bb6} (mass and angular momentum), and Schwartzschild \cite{Sch1916} (mass only) are limiting cases of the KN solution. Thus, for all $3+1$ black holes, the characteristics KN solution are important. Significantly, the event horizon, a topological two-sphere $(S^2)$ \cite{bb7} is the null surface of these black hole spacetimes. This surface acts as a one-way barrier between events occurring outside the horizon and the black hole's interior.

\indent The spacetime geometry of a black hole is explained by the spherically symmetric non-static line element (see \cite{SDHK_interior_2020})
\begin{equation}
ds^2 = -B(z, t)dt^2 + A(z, t)dz^2 + t^2\left(d\theta^2 + \sin^2 \theta d\phi^2\right),
\end{equation}
where Einstein's field equations for empty spacetime with vanishing Ricci tensor can be solved to provide the coefficient functions. 
 This is the empty annular area of spacetime for a spherical star inside its Schwarzschild radius $2m$ and outside its physical radius $a$ ($< 2m$).
The conclusion displays as
\begin{equation}\label{eq1.2}
ds^2 = -\left(\frac{2\xi}{t} - 1\right)^{-1}dt^2 + \left(\frac{2\xi}{t} - 1\right)dz^2 + t^2\left(d\theta^2 + \sin^2 \theta d\phi^2\right).
\end{equation}
The metric for an interior black hole is thus represented by (\ref{eq1.2}) \cite{{ROSA}, {HK}}, where $\xi$ can be computed via a direct confrontation with the exterior Schwarzschild solution. In the exterior region $z > a$ of a black hole the empty spacetime is represented by the metric (\ref{eq1.2}), which is the interior black hole solution. It is valid for $a$ such that $2m < z < a$. For further information, we refer the reader to \cite{ROSA} along with the associated references for insights into the cosmological significance and the physical relevance of the interior black hole solution. The nature of spacetime undergoes an interesting transformation in the interior black hole solution, though the external spatial radial and temporal coordinates change to temporal and spatial coordinates, respectively. As a result, the interior black hole solution is represented by a non-static spacetime because its metric coefficients depend on time.
We must mention that Shaikh et. al.\cite{SDHK_interior_2020} thorouhly studied the curvature properties of IBH spacetime and explored that such spacetime admits various types of pseudosymmetric condition and Roter type relation.

Assume that $(\mathcal{M}_{\nu}, g_{ab})$ is a $\nu$-dimensional semi-Riemannian manifold with the metric $g_{ab}$ and signature $(\mu, \nu)$ or $(\nu, \mu)$, where $\mu$ varies from $0$ to $\nu-1$ and $\mathcal{M}_{\nu}$ forms an open connected subset of $\mathbb{R}^\nu$. The semi-Riemannian manifold $(\mathcal{M}_{\nu}, g_{ab})$ is Lorentzian for $\mu \in \{1,\nu-1\} $ and Riemannian for $\mu = 0$. A $4$-dimensional connected Lorentzian manifold is called a spacetime. Let $\nabla,R,S$ and $\kappa$ be the Levi-Civita connection, Riemann Christoffel curvature , Ricci curvature and scalar curvature of $M_{\nu}$ respctively. Again in view of the energy-momentum tensor $T$ of rank $(0, 2)$ allows to understand the characteristics of the density and flux of energy as well as momentum, which enhance the perception of energy-momentum distribution due to the gravitational effects.
 The metric tensor, scalar curvature, Ricci curvature, Riemann curvature and energy-momentum tensor are essential to comprehend the shape of the spacetime, and they have significant impact on the geometry of underlying spacetime.

The geometric quantity ``curvature" and its qualities are therefore assumed to reveal a variety of physical features in space.
  For example, the curvature condition $R^{ab}_{pq}R_{abrs}=0$ indicates that a Brinkmann wave is a pp-wave  \cite{Brink1925, SBH21, MS2016, SG1986}, and the spacetime with pseudosymmetric Weyl tensors are of Petrov type D \cite{DHV2004}.
  A perfect fluid solution correlates to a quasi-Einstein spacetime, and vice versa, as shown in references \cite{C01, SKH11, SYH09}. We designate the perfect fluid Friedmann-Lemaître-Robertson-Walker (FLRW) \cite{DDHKS00} spacetime as quasi-Einstein, emphasizing its significance in cosmological models that describe an isotropic and homogeneous world, as well as its unique geometric characteristics.

However, the curvature tensors along with their covariant derivatives up to higher orders intricately expose the geometry of a semi-Riemannian manifold. There is symmetry at every point in the manifold $\mathcal{M}_{\nu}$ when all local geodesic symmetries are isometric, according to the condition $\nabla_a R_{pqrs} = 0$, which indicates a locally symmetric space \cite{Cart26}. This characteristic emphasizes the consistency of geometric properties across points that adjoin and serves as the foundation to differential geometry. This idea was developed further by Ruse \cite{Patt52, Ruse46, Ruse49a, Ruse49b}, who introduced ``kappa" spaces, which loosen this strict necessity to investigate larger classes of geometric structures. Such scaling-capable ``kappa" spaces were first identified as an important class of structures, originating from studies on maintaining curvature under parallel transport. This was referred to as recurrent manifolds by Walker \cite{Walk50}, and further studies have presented a number of families of manifolds that expand and improve upon the idea of recurrences. Those who wish to delve deeper into these lots of individuals will find thorough discussions in articles \cite{SAR13, SK14, SKA18, SP10, SR11, SRK15, SRK17, SRK18}. Formerly, Shirokov \cite{Shi25} studied spaces with covariantly constant tensors, further information can be found in related literature \cite{Shi98}.

The relation $R \cdot R = 0$ provides a concise expression for the integrability criteria resulting from the equations $\nabla R = 0$. These situations have been thoroughly examined and evaluated in the works of Cartan \cite{Cart46} and Shirokov \cite{Shi25} . The manifold $\mathcal{M}_{\nu}$ is defined as semisymmetric when $R \cdot R = 0$. Sinyukov's work \cite{Sin54} exhaustively examines the precise circumstances in which $R \cdot R = L_{R} Q(g, R)$, where $L_{R}$ represents any smooth function defined on the subset  
$\{x \in \mathcal{M} : Q(g, R) x \neq 0\}$ with $Q(g, R)$ representing the Tachibana tensor. The above facts are known as generalized symmetric and are described in detail in \cite{MIKES76} and \cite{Add1}, with more compact formulas found in \cite{MIKES88, MIKES92, MIKES96, MSV15, MVH09}.

Adamów and Deszcz \cite{AD83} developed a second order symmetry that goes beyond semisymmetry as a result of their in-depth study on completely umbilical hypersurfaces and their comprehensive investigation of symmetrical characteristics inside Einstein's field equations. In differential geometry and mathematical physics, this advancement significantly improves our understanding of geometric structures and their symmetries. The notion of pseudosymmetry is demonstrated by several types of spacetimes that serve as models of pseudosymmetric manifolds, including the Reissner-Nordstrom black hole \cite{Reis1916, Nord1918}, Schwarzschild \cite{Sch1916}, and Robertson-Walker spacetimes \cite{ADEHM14, DK99}. In \cite{HV07, HV07A, HV09}, Haesen and Verstraelen effectively explain the ideas of local symmetry, semisymmetry, and pseudosymmetry. The concept of Deszcz's  pseudosymmetry has been a major topic in cosmology and general relativity for the past forty years due to its numerous applications. In \cite{ADEHM14, DK99, Kowa06, SAA18, SAAC20, SDKC19, SAACD_LTB_2022, EDS_sultana_2022, SADM_TCEBW_2023, SAS_EiBI_2024, SASK_pgm_2024,SHDS_hayward, SHS_warped, SHS_Bardeen, SHSB_VBDS} a variety of spacetimes have been found to be exemplifying pseudosymmetry. These spacetimes continue to be essential in improving our knowledge of geometric and physical properties within theoretical frameworks. Furthermore, an extensive study on geodesic mappings across different symmetric Riemannian manifolds was carried out by Mikeš et al. \cite{MIKES76, MIKES88, MIKES92, MIKES96, MS94, MSV15, MVH09}. Remarkably, as stated in \cite{SDHJK15} the ideas of Deszcz and Chaki pseudosymmetries may resemble, adding to a more sophisticated comprehension of the geometric structures and features in spacetime analysis.
 
To investigate the evolution of Riemannian metrics on compact three-dimensional manifolds with positive Ricci curvature, Richard Hamilton created the notion of Ricci flow in 1982 \cite{Hamilton1982}. Plenty of investigations have been conducted on Ricci solitons, which are self similar solution of Ricci flow and also an improvement on the idea of Einstein metrics \cite{Bess87,Brink1925,SYH09,S09}. The idea of Ricci solitons has been extended over time to include a number of generalizations, such as almost Ricci solitons, $\eta$-Ricci solitons and almost $\eta$-Ricci solitons \cite{SMM2022,SMM2021,SM23,Absos23,SMM2024} etc.
 
A Ricci soliton in a semi-Riemannian manifold $\mathcal{M}_{\nu}$ is defined as 
$$
\frac{1}{2}\pounds_\zeta g + S - \phi_1 g = 0,
$$
where $\phi_1$ is a constant, and $\pounds_\zeta$ represents the Lie derivative in the direction of the smooth vector field $\zeta$. If $\phi_1 < 0$, the Ricci soliton is expanding; if $\phi_1 = 0$, it is steady; and if $\phi_1 > 0$, it is shrinking. If $\phi_1$ is a non-constant smooth function on $\mathcal{M}_{\nu}$, then  the soliton  is termed as an almost Ricci soliton \cite{Pigola2011}.

A Ricci soliton becomes an Einstein manifold if its soliton vector field $\zeta$ is Killing. Also, $\mathcal{M}_{\nu} $ is known as the $\eta$-Ricci soliton \cite{Cho2009} if there is a non-zero $1$-form $\eta$ on $\mathcal{M}_{\nu} $ with constants $\phi_1$ and $\phi_2$ such that
$$
\frac{1}{2}\pounds_\zeta g + S - \phi_1 g + \phi_2 (\eta \otimes \eta) = 0.
$$
If $\phi_1$ and $\phi_2$ are smooth functions on $\mathcal{M}_{\nu}$, then the $\eta$-Ricci soliton is referred to as an almost $\eta$-Ricci soliton \cite{Blaga2016}.

Alongside with Ricci flow, Hamilton \cite{Hamilton1988} presented the idea of Yamabe flow. Güler and Crășmăreanu \cite{Guler2019} developed the Ricci-Yamabe flow, a geometric flow that incorporates the feature of Yamabe and Ricci flow. Likewise Ricci-Yamabe solitons are the self-similar solutions of the Ricci-Yamabe flow, whereas Yamabe solitons are the self-similar solutions of the Yamabe flow. If the metric tensor $g$ and the Ricci curvature $S$ in a semi-Riemannian manifold $\mathcal{M}_{\nu} $ admits the equation
$$
\frac{1}{2}\pounds_\zeta g + \psi_1 S + \left(\phi_1 - \frac{1}{2}\psi_2 {\kappa}\right) g = 0,
$$
where $\psi_1$, $\psi_2$ and $\phi_1$ are constants and ${\kappa}$ is the scalar curvature then  $\mathcal{M}_{\nu} $ is called a Ricci-Yamabe soliton \cite{Siddiqi2020}. The study of geometric flows is developed by the recent discovery of Ricci-Yamabe solitons, which offers a cohesive framework that combines Yamabe and Ricci solitons, allowing for a greater understanding of the characteristics and behavior of manifolds under these combined flows. Moreover, by allowing $\psi_1$, $\psi_2$ and $\phi_1$ to   be non-constant smooth functions on $\mathcal{M}_{\nu} $, it characterizes an almost Ricci-Yamabe soliton \cite{Siddiqi2020}. Certain value of the constants $\psi_1$, $\psi_2$, and $\phi_1$ configures unique geometric structures. For instance the Yamabe soliton is obtained for the value $\psi_1 = 0$ and $\psi_2 = 2$, whereas a Ricci soliton is represented if $\psi_1 = 1$ and $\psi_2 = 0$.

Additionally, expanding the notion of Ricci-Yamabe soliton with a  non-zero 1-form $\eta$, an $\eta$-Ricci-Yamabe soliton \cite{Siddiqi2020} is defined as follows
$$
\frac{1}{2}\pounds_\zeta g + \psi_1 S + \left(\phi_1 - \frac{1}{2}\psi_2 {\kappa}\right) g + \phi_2 \eta \otimes \eta = 0,
$$
where $\phi_1$ is a constant. More valuable geometric structures and their consequences in differential geometry are demonstrated by exploring almost $\eta$-Ricci-Yamabe solitons. Several research articles on Ricci solitons, Yamabe solitons, and their generalizations have been published over the past decade (see, \cite{AliAhsan2013, AliAhsan2015, Ahsan2018, SDAA_LCS_2021, SMM2022, SM23,Absos23,SMB24} and references therein). These investigations have elevated the subject to the forefront of contemporary differential geometry research and development.
  
Shaikh and his collaborators \cite{SAA18,SKS19,SAACD_LTB_2022,SADM_TCEBW_2023,SHS_Bardeen,SAAC20,SAAC20N,SAS_EiBI_2024,SASK_pgm_2024,SBH21,SC21,SDAA_LCS_2021,ShaikhDatta2022,SDC,SDKC19,SHDS_hayward,SHS_warped,SHS_Bardeen,SHSB_VBDS,SK16srs,SKS19} have carried out significant studies on the curvature-constrained geometric structures in various spacetimes. Research on black holes examines how singularities offer an understanding of the information paradox and quantum gravity by demonstrating the behavior of spacetime under extreme curvature. Along with computational models, numerical relativity simulations like those done with the Einstein Toolkit are very important for studying high-energy physics situations and checking how stable other theories are, such as gravastars or fuzzballs. Hence, the IBH spacetime is a fascinating topic for study about its curvature-related aspects because it lacks a thorough explanation of its geometric characteristics in the literature of differential geometry. In particular, the purpose of this work is to provide insight into the curvature properties of the IBH spacetime. Recently Shaikh et.al. \cite{SDHK_interior_2020} studied the curvature properties of interior black hole spacetimes and obtained some pseudosymmetric nature of such spacetimes, yet the curvature related geometric properties of such spacetime are not fully determined. Through analysis of the article \cite{SDHK_interior_2020}, it has been observed that the IBH spacetimes do not admit the following structures: Ricci semisymmetric, quasi-Einstein, Codazzi type Ricci tensor, cyclic Ricci symmetric, Chaki pseudo symmetric, Chaki pseudo Ricci symmetric, weakly symmetric, weakly Ricci symmetric, hyper generalized recurrent, weakly generalized recurrent, quasi generalized recurrent etc. Motivated by the study on \cite{SDHK_interior_2020}, in present article, we have explored the geometric properties of IBH spacetime and obtained some deeper curvature related geometric properties of such spacetime. The most interesting fact that the IBH spacetime neither admits Ricci soliton nor Yamabe soliton with respect to the non-Killing vector fields    $\frac{\partial}{\partial t}$ and $\frac{\partial}{\partial \theta}$. Also, IBH spacetime does not realize curvature collineation and Ricci collineation. However, IBH spacetime admits generalized curvature inheritance with respect to the Riemann, conformal, concircular, conharmonic and Weyl projective curvature. Moreover, it is revealed that its conformal $2$-forms are recurrent, $2$-quasi-Einstein, Einstein manifolds of level $2$, Roter type etc. Analysis using the Tachibana tensor associated with the energy-momentum tensor $T$ demonstrates that the spacetime satisfies various types of pseudosymmetric structure.  The study concludes with a thorough comparison of the geometric characteristics of the IBH spacetime with the Kiselev black hole (briefly, KBH) spacetime.

The arrangement of the article is as follows: In Section $2,$ we define some important geometric structures that are necessary for our study. A complete analysis of the IBH spacetime is presented in Section $3$, from which we obtain and discuss a number of interesting conclusions. The geometric features related to the energy-momentum tensor of the IBH spacetime are examined in Section $4$. Section $5$ is enthusiastic about investigating curvature inheritance and collineation presented in the IBH spacetime. Finally, in Section $6$, we compare the IBH spacetime to the KBH spacetime, focusing on their respective geometric structures.
\vspace{-4 mm}
\section{\bf Curvature restricted geometric properties}
For any two symmetric tensors $\mathcal{H}$ and $\mathcal{F}$ of type $(0,2)$, the Kulkarni-Nomizu product is a $(0,4)$ tensor, which is represented as $(\mathcal{H} \wedge \mathcal{F}),$  and is defined as (see, \cite{DGHS11,DHJKS14, Glog02, G08, Kowa06}):
$$(\mathcal{H}\wedge \mathcal{F})_{cdef}=2\mathcal{H}_{c[f}\mathcal{F}_{e]d} +2\mathcal{H}_{d[e}\mathcal{F}_{f]c},$$
where the notation $[.]$ indicates the antisymmetrization of the index pairs, guaranteeing antisymmetry concerning those indices.

The $(1,3)$ Riemann, conharmonic, concircular, conformal, and projective curvatures are defined in the following ways:
\bea
R^p_{qrs}&=& 2\left(\Gamma^u_{q[r}\Gamma^p_{s]u} + \partial_{[s}\Gamma^p_{r]q}\right),\nonumber\\
K^p_{qrs}&=& R^p_{qrs} - \frac{2}{n-2}\left( S^p_{[q}g_{r]s} + \delta^p_{[q}S_{r]s}\right),\nonumber \\
W^p_{qrs}&=& R^p_{qrs} - {\kappa}\frac{2}{n(n-1)}\delta^p_{[q}g_{r]s}, \nonumber \\
C^p_{qrs}&=& R^p_{qrs}+\frac{2}{n-2}\left( \delta^p_{[q}S_{r]s} +S^p_{[q}g_{r]s}\right) -{\kappa}\frac{2}{(n-1)(n-2)}\delta^p_{[q}g_{r]s} ,\nonumber \\
P^p_{qrs}&=& R^p_{qrs} -\frac{2}{n-1}\delta^p_{[q}S_{r]s}, \nonumber
\eea
where $\Gamma^q_{rs}$ denote the coefficients of the connection, $S^q_r$  denotes the $(1,1)$ Ricci curvature and $\partial_p = \frac{\partial}{\partial x^p}.$

The $(0,4)$ tensors $R_{pqrs}$, $K_{pqrs}$, $W_{pqrs}$, $C_{pqrs}$, and $P_{pqrs}$ are generated by lowering the indices through the metric $g_{pq}$. They are represented as follows:
\bea
R_{pqrs}&=& g_{p\alpha}(\partial_s \Gamma^\alpha_{qr}-\partial_r \Gamma^\alpha_{qr}+\Gamma^\beta_{qr}\Gamma^\alpha_{\beta s}-\Gamma^\beta_{qs}\Gamma^\alpha_{\beta r}) , \nonumber \\
K_{pqrs}&=& R_{pqrs} - \frac{1}{n-2} (g\wedge S)_{pqrs} , \nonumber \\
W_{pqrs}&=& R_{pqrs} - \frac{{\kappa}}{2n(n-1)} (g\wedge g)_{pqrs}, \nonumber \\
C_{pqrs}&=& R_{pqrs}-\frac{1}{n-2}(g\wedge S)_{pqrs}+\frac{\kappa}{2(n-1)(n-2)}(g\wedge g)_{pqrs} ,\nonumber \\
P_{pqrs}&=& R_{pqrs} -\frac{1}{n-1}(g_{ps}S_{qr}-g_{qs}S_{pr}). \nonumber
\eea

For a $(0,l),$ $(l \geq 1)$ tensor $U$, the $(0,l+2)$ tensors $(A \cdot U)$ (see, \cite{DG02, DGHS98, DH03, SDHJK15, SK14}) and $Q(\mathscr{P},U)$ (see, \cite{DGPSS11, SDHJK15, SK14, Tach74}) are given as follows:

\beb
(A\cdot U)_{b_1b_2\cdots b_lws}&=&-\left[ A^\alpha_{wsb_1}U_{\alpha b_1\cdots b_l}+ \cdots + A^\alpha_{wsb_l}U_{b_1\cdots \alpha}\right], \\
Q(\mathscr{P},U)_{b_1b_2\cdots b_l uv}&=&\mathscr{P}_{vb_1}U_{ub_2\cdots b_l}+ \cdots + \mathscr{P}_{ub_l}U_{b_1b_2\cdots v}\\ 
&-& \mathscr{P}_{ub_1}U_{vb_2\cdots b_l}- \cdots - \mathscr{P}_{ub_l}U_{b_1b_2\cdots v},
\eeb
where $\mathscr{P}_{ij}$ is a $(0,2)$ symmetric tensor and $U^i_{jkl}$ is a $(1,3)$ tensor.

\begin{defi} (\cite{AD83, Cart46,  Desz92, Desz93, DGHZ15, DGHZ16,SAAC20N, SK14, SKppsnw, Szab82, Szab84, Szab85})

If the tensor $A \cdot U$ is linearly dependent on $Q(g, U),$ given as $A \cdot U = f_U Q(g, U)$, where $f_U$ is a smooth function on $\mathcal{M}_{\nu}$, then the manifold $\mathcal{M}_{\nu}$ is called $U$-pseudosymmetric due to $A$. Furthermore, if the condition $A \cdot U = \bar{f}_U Q(S, U)$ holds on $\mathcal{M}_{\nu}$, where  $\bar{f}_U$ is another smooth function on $\mathcal{M}_{\nu}$, then the manifold is mentioned as Ricci generalized $U$-pseudosymmetric due to $A$.
\end{defi}

The manifold has pseudosymmetric and Ricci generalized pseudosymmetric features when $A = R$ and $U = R$. It is possible to identify several kinds of pseudosymmetric and Ricci generalized pseudosymmetric manifolds by investigating different curvatures, namely Riemann, Ricci and other curvature tensors.

When $A \cdot U = 0$, a manifold is said to be $U$-semisymmetric due to $A$. In particular, a semisymmetric manifold fulfills the requirement $R \cdot R = 0$ \cite{Cart46, Szab82, Szab84, Szab85}. Naturally, semisymmetric manifolds are a subset of pseudosymmetric manifolds, highlighting the broader implications of specific curvature conditions. In differential geometry, this variation emphasizes how curvature criteria, like $A \cdot U = 0$, define several symmetric properties.

\begin{defi}$($\cite{C01,DGHZ16, DGJZ-2016, DGP-TV-2011, S09, SKH11, SK19,SYH09}$)$ 
 
If the tensor $(S - \mathscr{X} g)$ has rank $k$, then $\mathcal{M}_{\nu}$ is called a $k$-quasi-Einstein manifold for a scalar $\mathscr{X}$ and $0 \leq k \leq (\nu-1)$. In particular, it is called a quasi-Einstein \cite{SKH11,SYH09} (resp., Einstein) manifold when $k = 1$ (resp., $k = 0$). 
A manifold is classified as Ricci simple when it is a quasi-Einstein manifold with $\mathscr{X}=0$.

\end{defi}

For instance, Robertson-Walker spacetime \cite{ARS95, Neill83, SKMHH03} is classified as quasi-Einstein, Kaigorodov spacetime \cite{SDKC19} is Einstein, Kantowski-Sachs spacetime \cite{SC21} and Som-Raychaudhuri spacetime \cite{SK16srs} are examples of 2-quasi-Einstein manifolds, and Vaidya spacetime \cite{SKS19}, Gödel spacetime \cite{DHJKS14} and Morris-Thorne spacetime \cite{ECS22} are Ricci simple manifolds.

\begin{defi}
A manifold $\mathcal{M}_{\nu}$ describes cyclic parallel Ricci tensor (see \cite{Gray78, SB08, SJ06, SJ07}), if it fulfills the relation
\[ (\nabla_{\mathscr{V}_1} S)(\mathscr{V}_2, \mathscr{V}_3) + (\nabla_{\mathscr{V}_2} S)(\mathscr{V}_3, \mathscr{V}_1) + (\nabla_{\mathscr{V}_3} S)(\mathscr{V}_1, \mathscr{V}_2) = 0. \]
 Moreover, a Codazzi Ricci tensor (see \cite{F81, S81}) is defined by the condition
\[ (\nabla_{\mathscr{V}_1} S)(\mathscr{V}_2, \mathscr{V}_3) = (\nabla_{\mathscr{V}_2} S)(\mathscr{V}_1, \mathscr{V}_3) \]
for smooth vector fields $\mathscr{V}_1$, $\mathscr{V}_2$, and $\mathscr{V}_3$ on $\mathcal{M}_{\nu}.$
\end{defi}
Notably, the Ricci tensor of $(t-z)$-type plane wave spacetime \cite{EC21} is categorized as Codazzi type, but the Gödel spacetime \cite{DHJKS14} is shown as cyclic parallelism.

\begin{defi} $($\cite{Bess87, SK14, SK19}$)$ 
If the relation $\epsilon_1 S^4 + \epsilon_2 S^3 + \epsilon_3 S^2 + \epsilon_4 S + \epsilon_5 g = 0, \quad (\epsilon_1 \neq 0),$  holds on a manifold $\mathcal{M}_{\nu},$ where $\epsilon_i$ are smooth functions on $\mathcal{M}_{\nu},$ then it is called an Einstein manifold of level $4$. If $\epsilon_1 = 0$ but $\epsilon_2 \neq 0$ (or $\epsilon_1 = \epsilon_2 = 0$ but $\epsilon_3 \neq 0)$, then  $\mathcal{M}_{\nu}$ is an Einstein manifold of level $3$ (or level $2,$ respectively).
\end{defi}

It is observed that the Vaidya-Bonner spacetime \cite{SDC} and Lifshitz spacetime \cite{SSC19} are examples of Einstein manifolds with level $3,$ whereas the Siklos spacetime \cite{SDKC19} and Nariai spacetime \cite{SAAC20N} are examples of Einstein manifolds with level $2.$
\begin{defi} 
The Riemann curvature tensor of a generalized Roter type manifold $\mathcal{M}_{\nu}$ can be expressed as (\cite{Desz03, DGJPZ13, DGJZ-2016, DGP-TV-2015, SK16,SK19}):
$$ R = (\mathcal{B}_{11} S^2 + \mathcal{B}_{12} S + \mathcal{B}_{13} g) \wedge S^2 + (\mathcal{B}_{22} S + \mathcal{B}_{23} g) \wedge S + \mathcal{B}_{33} (g \wedge g)$$

here $\mathcal{B}_{ij}$ are scalar coefficients. 
The manifold is categorized as a Roter type manifold if these coefficients minimize the linear dependency to combinations involving $R$, $g \wedge g$,  $g \wedge S$ and $S \wedge S$ (\cite{Desz03, DG02, DGP-TV-2011, DPSch-2013, Glog-2007}).
\end{defi}
Significantly, Nariai spacetime \cite{SAAC20N}, Melvin magnetic spacetime \cite{SAAC20} and Robinson-Trautman spacetime \cite{SAA18} are classified as Roter type manifolds. However, Lifshitz spacetime \cite{SSC19} and Vaidya-Bonner spacetime \cite{SDC} belong to generalized Roter type manifolds.
\begin{defi} \cite{TB89, TB93}
A weakly $A$-symmetric manifold $\mathcal{M}_{\nu}$ is represented by the equation as follows:
\beb
	(\nabla_{\mathcal{T}} A)(\mathcal{T}_1,\mathcal{T}_2,\mathcal{T}_3,\mathcal{T}_4)&=& \Pi(\mathcal{T})\otimes A(\mathcal{T}_1,\mathcal{T}_2,\mathcal{T}_3,\mathcal{T}_4)+ \omega_1(\mathcal{T}_1)\otimes A(\mathcal{T},\mathcal{T}_2,\mathcal{T}_3,\mathcal{T}_4)\\ &+& \omega_1(\mathcal{T}_2)\otimes A(\mathcal{T}_1,\mathcal{T},\mathcal{T}_3,\mathcal{T}_4)+ \omega_2(\mathcal{T}_3)\otimes A(\mathcal{T}_1,\mathcal{T}_2,\mathcal{T},\mathcal{T}_4)\\ &+& \omega_2(\mathcal{T}_4)\otimes A(\mathcal{T}_1,\mathcal{T}_2,\mathcal{T}_3,\mathcal{T}),
	\eeb
	where  $\Pi$, $\omega_1$ and $\omega_2$  are $1$-forms on $\mathcal{M}_{\nu}.$
	In particular, the manifold $\mathcal{M}_{\nu}$ reduces to a recurrent \cite{Ruse46,Ruse49a,Walk50} if  $\omega_1$ = $\omega_2$ = $0$. On the other hand, 
	$\mathcal{M}_{\nu}$ becomes a Chaki pseudosymmetric manifold \cite{Chak87, Chak88} if $\omega_1$ = $\omega_2 = \Pi\slash2$.

\end{defi}

\begin{defi}$($\cite{DD91,DGJPZ13, MM12a, MM12b, MM13,MM22b}$)$
Let $I$ be a $(0,4)$ tensor field. A symmetric $(0, 2)$-tensor $\beta$ is $I$-compatible if
\beb
	\mathop{\mathcal{S}}_{\mathcal{T}_1,\mathcal{T}_2,\mathcal{T}_3} A(\mathcal{I}\mathcal{T}_1,\mathcal{T},\mathcal{T}_2,\mathcal{T}_3)=0,
	\eeb
holds on $\mathcal{M}_{\nu},$ where $\mathcal{I}_{\beta}$ is the endomorphism corresponding to $\beta.$ Further, an $1$-form $\Upsilon$ is said to be $I$-compatible if $\Upsilon \otimes \Upsilon$
is $I$-compatible.

\end{defi}

When considering the curvature tensors $R$ (Riemann curvature tensor), $C$ (conformal curvature tensor), $W$ (concircular curvature tensor), $P$ (projective curvature tensor), or $K$ (conharmonic curvature tensor), we obtain their respective compatibility conditions by replacing $I$ with the appropriate tensor.

\begin{defi}\label{defi2.8} 
If the curvature 2-forms $\omega_{(I)}^m l$ \cite{LR89} satisfy the relation
\beb
	\mathop{\mathcal{S}}_{\mathcal{T}_1,\mathcal{T}_2,\mathcal{T}_3}(\nabla_{\mathcal{T}_1}I)(\mathcal{T}_2,\mathcal{T}_3,\mathcal{T}_4,\mathcal{T}_5)=\mathop{\mathcal{S}}_{\mathcal{T}_1,\mathcal{T}_2,\mathcal{T}_3}\Sigma(\mathcal{T}_1)I(\mathcal{T}_2,\mathcal{T}_3,\mathcal{T}_4,\mathcal{T}_5)
	\eeb
for an $1$-form $\Sigma$, then they are referred to as recurrent \cite{MS12a,MS13a,MS14} and vice-versa.
Moreover, the 1-forms $\Lambda_{(T)l}$ are considered as recurrent \cite{SKP03} if they satisfy the equation
\[
(\nabla_{\mathcal{T}_1}T)(\mathcal{T}_2,\mathcal{T}) - (\nabla_{\mathcal{T}_2}T)(\mathcal{T}_1,\mathcal{T}) = \Gamma(\mathcal{T}_1) T(\mathcal{T}_2,\mathcal{T}) - \Gamma(\mathcal{T}_2) T(\mathcal{T}_1,\mathcal{T})
\]

on  $\mathcal{M}_{\nu},$ where $\Gamma$ is an $1$-form.

\end{defi}

\begin{defi}$($\cite{P95, Venz85}$)$ The set of solutions to the equation
 	\beb
	\mathop{\mathcal{S}}_{\mathcal{T}_1,\mathcal{T}_2,\mathcal{T}_3}\Gamma(\mathcal{T}_1)\otimes A(\mathcal{T}_2,\mathcal{T}_3,\mathcal{T}_4,\mathcal{T}_5)=0.
	\eeb
forms a vector space $L(\mathcal{M}_{\nu}),$  here $\mathcal{S}$ denotes a summation over $\mathcal{T}_1,$ $\mathcal{T}_2$ and $\mathcal{T}_3.$  According to Venzi's terminology, if the dimension of $L(\mathcal{M}_{\nu}) \geq 1,$ then $\mathcal{M}_{\nu}$ is known as $A$-space .

	\end{defi}
%

\begin{defi}The manifold $\mathcal{M}_{\nu}$ has a symmetry associated with $\zeta$ if $\pounds_\zeta g = 0$, and $\zeta$ is referred to as a Killing vector field in this context.
\end{defi}
Katzin et al. introduced the concept of curvature collineation for the $(1,3)$-type curvature tensor\cite{KLD1969,KLD1970} in $1969.$ It was accomplished by vanishing the Lie derivative of the Riemann curvature tensor of the $(1,3)$ type with respect to a smooth vector field. To elaborate on this concept, in $1992,$ Duggal \cite{Duggal1992} extending the concept to curvature inheritance for the $(1,3)$-type curvature tensor. During the past three decades, a considerable amount of articles (see, \cite{Ahsan1978, Ahsan1977_231, Ahsan1977_1055, Ahsan1987, Ahsan1995, Ahsan1996, Ahasan2005, AhsanAli2014, AA2012, AH1980, AliAhsan2012, SASZ2022, ShaikhDatta2022}) have been published in the literature, examining these types of symmetries.

\begin{defi}\label{def_CI} (\cite{Duggal1992}) A semi-Riemannian manifold $\mathcal{M}_{\nu}$ is referred to be curvature inheritance for a curvature tensor of type $(1,3)$ with a non-Killing vector field $\zeta$ and a scalar function $\mathcal{J}$ if it satisfies the equation $$\pounds_\zeta \widetilde{R} = \mathcal{J} \widetilde{R}.$$
The relationship between the $(1,3)$ curvature tensor $\widetilde{R}$ and the $(0,4)$ curvature tensor $R$ is given by  $$R(\mathcal{T}_1, \mathcal{T}_2, \mathcal{T}_3, \mathcal{T}_4) = g(\widetilde{R}(\mathcal{T}_1, \mathcal{T}_2) \mathcal{T}_3, \mathcal{T}_4).$$ Additionally, curvature collineation occurs when $\mathcal{J} = 0$ (i.e., $\pounds_\zeta \widetilde{R} = 0$) \cite{KLD1969, KLD1970}.




\end{defi}

 \begin{defi} \label{def_RI}(\cite{Duggal1992})
A semi-Riemannian manifold $\mathcal{M}_{\nu}$ reveals Ricci inheritance if there exists a vector field $\zeta$ and a scalar function  $\mathcal{J}$ such that

$$\pounds_\zeta S = \mathcal{J} S,$$

where $S$  denotes the Ricci tensor of the manifold  $\mathcal{M}_{\nu}.$

Specifically, this condition corresponds to Ricci collineation, which states that $\pounds_\zeta S = 0$, when $\mathcal{J} = 0$.
\end{defi}
\begin{defi}[\cite{ShaikhDatta2022}]
A semi-Riemannian manifold $\mathcal{M}_{\nu}$ realizes generalized Ricci inheritance if there exists a vector field $\zeta$ and scalar functions  $\mathcal{J}_1$, $\mathcal{J}_2$ such that

$$\pounds_\zeta S = \mathcal{J}_1 S+\mathcal{J}_2g,$$
In particular:\\
$(i)$ If $\mathcal{J}_2=0,$ then $\mathcal{M}_{\nu}$ satisfies Ricci inheritance.\\
$(ii)$ If $\mathcal{J}_1$ and $\mathcal{J}_2,$ both vanishes then this condition reduces to Ricci collineation.
\end{defi}
For the $(0,4)$-type curvature tensor $R$, Shaikh and Datta \cite{ShaikhDatta2022} have presented a unique extension of curvature inheritance. This sophisticated idea is called generalized curvature inheritance, and its definition is as follows:
\begin{defi}\label{def_GCI} (\cite{ShaikhDatta2022})
A semi-Riemannian manifold $\mathcal{M}_{\nu}$ is referred to as generalized curvature inheritance for the $(0,4)$-type curvature tensor $R$ if there exists a non-Killing vector field $\zeta$ satisfying

$$\pounds_\zeta R = \mathcal{J} R + \mathcal{J}_1 g \wedge g + \mathcal{J}_2 g \wedge S + \mathcal{J}_3 S \wedge S,$$

where $ \mathcal{J}, \mathcal{J}_1, \mathcal{J}_2$ and  $\mathcal{J}_3 $ are scalar functions. 

In particular:\\
$(i)$ If $ \mathcal{J}_i = 0 $ for $ i = 1, 2, 3 $, then  $\mathcal{M}_{\nu}$  realizes curvature inheritance for the $(0,4)$-type curvature tensor $R$.\\
$(ii)$ If $\mathcal{J} = 0$ and $\mathcal{J}_i = 0$ for $ i = 1, 2, 3,$ then this condition minimizes to curvature collineation for the $(0,4)$-type curvature tensor $R$.
\end{defi}
Shaikh et al. \cite{ShaikhDatta2022} recently demonstrated that, the concepts of curvature inheritance given in Definition \ref{def_CI} (for (1,3)-type curvature tensor) and Definition \ref{def_GCI} (for (0,4)-type curvature tensor) are not equivalent for point-like global monopole spacetime \cite{ShaikhDatta2022}. 
similarly we can define the generalized curvature inheritance for other curvature tensors such as conformal, conharmonic,concircular etc.
\section{\bf Interior black hole spacetime admitting geometric structures} 
In Eddington-Finkelstein coordinates  $(t,z,\theta,\phi,)$, the metric tensor  of IBH spacetime is given by
$$g=\left(\begin{array}{cccc}
	 -\left(\frac{2\xi}{t} - 1\right)^{-1} & 0 & 0 & 0 \\
	0 &  \left(\frac{2\xi}{t} - 1\right) & 0 & 0 \\
	0 & 0 & t^2 & 0 \\
	0 & 0 & 0 & t^2\sin^2\theta 
\end{array}
\right).
$$
That is,
 \begin{eqnarray}
g_{11}= -\left(\frac{2\xi}{t} - 1\right)^{-1},\; g_{22}= -\left(\frac{2\xi}{t} - 1\right), \ \ g_{33}=t^2, \;g_{44}=t^2\sin^2\theta \text{ and } g_{ab}=0, otherwise. \; \label{1}
 \end{eqnarray}

Throughout the paper, we will use the following expressions:
$V_1=\xi-t \dot\xi ,$ $V_2=t-2 \xi,$ 
$V_3=t (t \ddot{\xi}-3 \dot{\xi})+3 \xi,$ $V_4=t \ddot{\xi}-2 \dot\xi,$ $ V_5=t (t \ddot{\xi}-4 \dot{\xi})+6 \xi,$ $V_6=t (2 t \ddot{\xi}-7 \dot{\xi})+9 \xi,$ 
$V_7=t (t \ddot{\xi}-5 \dot{\xi})+9 \xi,$ $V_8=t \dot{\xi}-3 \xi,$, $V_9=t^2 \ddot{\xi}-2 \xi,$ $V_{10}=t-4\xi,$ $V_{11}=t\ddot{\xi}+2\xi$ and $V_{12}=t\dot{\xi}+3\xi.$

where $\dot{\xi}$ denotes the differentiation of $\xi$ with respect to $t$
and $\ddot{\xi}$  is the second order differentiation of $\xi$ with respect to $t$.

As a result of the subsequent computation, we obtain the non-zero components of the Christoffel symbols of the second kind ($\Gamma^h_{ij}$), which are given as
 \begin{eqnarray}\label{2}
 \begin{cases}

 	\Gamma^{1}_{11} = -\frac{V_1}{t V_2} = -\Gamma^{2}_{12}, \ \ 
 	\Gamma^{3}_{13} = \frac{1}{t}= \Gamma^{4}_{14}, \ \ \Gamma^{1}_{22} = \frac{V_2 V_1}{t^3},$$ \\
 	$$\Gamma^{1}_{33} = -V_2 , \ \ \Gamma^{4}_{34} = \cot \theta, \ \ \Gamma^{1}_{44} = -V_2 \sin^2\theta, \\ 
 	\Gamma^{3}_{44} = -\sin \theta \cos \theta$$.\\
\end{cases}
  \end{eqnarray}

Again, the non-zero terms of the Riemann curvature tensor and Ricci tensor are calculated as follows:
\begin{eqnarray}\label{3}
 \begin{cases}

 	$$ R_{1212} = -\frac{t^2 \ddot{\xi}-2 t \dot{\xi}+2 \xi}{t^3}, \ \ \ R_{1313} = -\frac{V_1}{V_2},
 	R_{1414} = -\frac{V_1\sin^2\theta}{V_2},$$\\
 	$$ R_{2323} = \frac{V_2 V_1}{t^2}=\frac{1}{\sin^2\theta} R_{2424}, \ \ \ 
 	R_{3434} = 2 t \xi \sin^2\theta.$$
 \end{cases}
  \end{eqnarray}
 
 \begin{eqnarray}\label{4}
 \begin{cases}
 	S_{11} = -\frac{\ddot{\xi}}{t-2 \xi}       = - \frac{ \ddot{\xi} }{t} g_{11},\ \ \
 	S_{22} = \frac{(t-2 \xi) \ddot{\xi}}{t^2}  = - \frac{ \ddot{\xi} }{t} g_{22},\\
 	S_{33} = -2 \dot{\xi} =  - \frac{2 \dot{\xi}}{t^{2}} g_{33},\ \ \ 
 	S_{44} = -2 \dot{\xi} \sin^{2} \theta  =  - \frac{2 \dot{\xi}}{t^{2}} g_{44}.\ \ \
  \end{cases}
   \end{eqnarray}

  Finally, the scalar curvature is given by
  \begin{equation}
      {\kappa}=-\frac{2(2\dot{\xi} + t\ddot{\xi})}{t^2}.\label{5}
  \end{equation}
  
%


Let $\mathscr{Z}^{1} = R \cdot R,$ $\mathscr{Z}^{2} =\nabla R,$ $\mathcal{K}^1 = Q(g, R)$ and $\mathcal{K}^2 = Q(S, R).$ We now provide the non-zero components of $\mathscr{Z}^{1},$ $\mathscr{Z}^{2},$ $\mathcal{K}^1$ and $\mathcal{K}^2,$ considering their symmetries, as follows:

\begin{eqnarray}\label{RR}
\begin{cases}
$$\mathscr{Z}^{1}_{1223,13}=-\frac{V_1 V_3}{t^4} = -\mathscr{Z}^{1}_{1213,23}=\frac{1}{\sin^2\theta}\mathscr{Z}^{1}_{1224,14}=\frac{1}{\sin^2\theta}\mathscr{Z}^{1}_{1214,24},$$\\
$$\mathscr{Z}^{1}_{1434,13}=-\frac{V_2 V_8\sin^2\theta}{t V_2} = -\mathscr{Z}^{1}_{1334,14},$$\
$$\mathscr{Z}^{1}_{243423}=-\frac{(t-2 \xi) \sin^2\theta (t \dot{\xi}-3 \xi) (t \dot{\xi}-\xi)}{t^3} = -\mathscr{Z}^{1}_{233424};$$
	\end{cases}
	\end{eqnarray}

\begin{eqnarray}\label{R}
\begin{cases}
$$ \mathscr{Z}^{2}_{1212,1}=\frac{- t^3 \xi^3 +3 t^2 \ddot{\xi}-6 t \dot{\xi}+6 \xi}{t^4}, \mathscr{Z}^{2}_{1223,3}=\frac{V_2 V_3}{t^3} = - \mathscr{Z}^{2}_{2323,1}=\frac{1}{\sin^2\theta}\mathscr{Z}^{2}_{1224,4}=-\frac{1}{\sin^2\theta} \mathscr{Z}^{2}_{2424,1},$$ \\
 \mathscr{Z}^{2}_{1313,1}=\frac{V_3}{t V_2}=\frac{1}{\sin^2\theta}\mathscr{Z}^{2}_{1414,1},$$\
$$ \mathscr{Z}^{2}_{1334,4}=\sin^2\theta (3 \xi-t \dot{\xi}) = - \mathscr{Z}^{2}_{1434,3} = -\frac{1}{2}\mathscr{Z}^{2}_{3434,1};$$
	\end{cases}
	\end{eqnarray}

\begin{eqnarray}\label{Q(g,R)}
\begin{cases}	
$$\mathcal{K}^1_{1223,13}=-\frac{V_3}{t} = -\mathcal{K}^1_{1213,23}=\frac{1}{\sin^2\theta}\mathcal{K}^1_{1214,24}=\frac{1}{\sin^2\theta}- \mathcal{K}^1_{1224,14},$$ \\
$$\mathcal{K}^1_{1434,13}=\frac{t^2 V_8 \sin^2\theta }{tV_2} = -\mathcal{K}^1_{1334,14},$$ \
$$\mathcal{K}^1_{2434,23}=V_2 V_8 \sin^2\theta =- \mathcal{K}^1_{2334,24};$$
	\end{cases}
		\end{eqnarray}

\begin{eqnarray}\label{Q(Ric,R)}
\begin{cases}	
	\mathcal{K}^2_{1223,13}=\frac{4 \dot{\xi}V_1+t (t \dot{\xi}+\xi)\ddot{\xi}}{t^3}=\frac{1}{\sin^2\theta}\mathcal{K}^2_{1224,14}=-\mathcal{K}^2_{1213,23}=\frac{1}{\sin^2\theta}\mathcal{K}^2_{1214,24}, \\ 
	\mathcal{K}^2_{1434,13}=-\frac{2 \sin^2\theta \left(\xi (t \ddot{\xi}+\dot{\xi})-t \dot{\xi}^2\right)}{V_2}=-\mathcal{K}^2_{1334,14}, \\

	\mathcal{K}^2_{2434,23}=\frac{2 V_2 \sin^2\theta \left(\xi (t \ddot{\xi}+\dot{\xi})-t \dot{\xi}^2\right)}{t^2}=-\mathcal{K}^2_{2334,24}.
	\end{cases}
		\end{eqnarray}




%
%

The conformal curvature tensor $C_{pqrs},$ considering its symmetry, has non-zero components and are presented as follows:

$$\begin{array}{c}
	C_{1212}=-\frac{V_5}{3 t^3},\
	C_{1313}=-\frac{V_ 5}{6 V_2}=\frac{1}{\sin^2\theta}C_{1414}, \\
	C_{2323}=\frac{V_5}{6}=\frac{1}{\sin^2\theta}C_{2424}, \ \ 
C_{3434}=\frac{-tV_5 \sin^2\theta}{3}.\\
\end{array}$$

The components of covariant derivatives of conformal curvature tensors which are not vanish, given as 
$$\begin{array}{c}
C_{1212,1}=\frac{18\xi-14t\dot{\xi}+5t^2\ddot{\xi}-t^3\xi^3}{3t^4}\ \
=2t^3V_2C_{1313,1}=\frac{2t^3V_2}{\sin^2\theta}C_{1414,1}\\
=-\frac{2t}{V_2}C_{2323,1}=\frac{2t}{V_2\sin^2\theta}C_{2424,1}=\frac{t^4}{\sin^2\theta}C_{3434,1},\\
C_{1223,3}=\frac{V_2V_5}{2t^3}=\frac{1}{\sin^2\theta}C_{1224,4}=\frac{t^3}{\sin^2\theta}C_{1334,4}=\frac{t^3}{\sin^2\theta}C_{1434,3}.
\end{array}$$

Let  $\mathscr{Z}^{3}=R\cdot C$, $\mathscr{Z}^{4}=C\cdot R$,  $\mathcal{K}^3=Q(g,C)$ and $\mathcal{K}^4=Q(S,C)$.  We now provide the non-zero components of $\mathscr{Z}^{3}$, $\mathscr{Z}^{4}$,  $\mathcal{K}^3$ and $\mathcal{K}^4$ considering their symmetries, as follows:

\begin{eqnarray}\label{RC}
\begin{cases}
		\mathscr{Z}^3_{1223,13}=-\frac{V_1V_5}{2 t^4}=-\frac{1}{\sin^2\theta}\mathscr{Z}^3_{1224,14}=-\mathscr{Z}^3_{1213,23}=-\frac{1}{\sin^2\theta}\mathscr{Z}^3_{1214,24}, \\ 
		\mathscr{Z}^3_{1434,13}=\frac{V_1V_5\sin^2\theta}{2tV_2}=-\mathscr{Z}^3_{1334,14},\\
		\mathscr{Z}^3_{2434,23}=-\frac{V_1V_2V_5\sin^2\theta}{2t^3}=-\mathscr{Z}^3_{2334,24}; \\
	
	\end{cases}
			\end{eqnarray}

\begin{eqnarray}\label{CR}
\begin{cases}
	\mathscr{Z}^4_{1223,13}=-\frac{V_3V_5}{6t^4}=\frac{1}{\sin^2\theta}\mathscr{Z}^4_{1224,14}=-\mathscr{Z}^4_{1213,23}=-\frac{1}{\sin^2\theta}\mathscr{Z}^4_{1214,24}, \\
	\mathscr{Z}^4_{1434,13}=-\frac{V_8V_5\sin^2\theta}{6tV_2}=-\mathscr{Z}^4_{1334,14}, \\
	\mathscr{Z}^4_{2434,23}=\frac{V_2V_8V_5\sin^2\theta}{6t^3}=-\mathscr{Z}^4_{2334,24}; 
	\end{cases}
			\end{eqnarray}

		
\begin{eqnarray}\label{Q(g,C)}
\begin{cases}
\mathcal{K}^3_{1223,13}=-\frac{V_5}{2t}=\frac{1}{\sin^2\theta}\mathcal{K}^3_{1224,14}=-\mathcal{K}^3_{1213,23}=-\frac{1}{\sin^2\theta}\mathcal{K}^3_{1214,24},\\	
	\mathcal{K}^3_{1434,13}=\frac{t^2\sin^2\theta V_5}{2V_2}=-\mathcal{K}^3_{1334,14}, \\
	\mathcal{K}^3_{2434,23}=-\frac{V_2V_5\sin^2\theta}{2}=-\mathcal{K}^3_{2334,24}; \\
\end{cases}
		\end{eqnarray}

\begin{eqnarray}\label{Q(Ric,C)}
\begin{cases}
\mathcal{K}^4_{1223,13}=\frac{\left(t \ddot{\xi}+4 \dot{\xi}\right) V_5}{6t^3}=\frac{1}{\sin^2\theta}\mathcal{K}^4_{1224,14}=-\mathcal{K}^4_{1213,23}=\frac{1}{\sin^2\theta}\mathcal{K}^4_{1214,24}, \\ 
	\mathcal{K}^4_{1434,13}=\frac{\sin^2\theta \left(t \ddot{\xi}+\dot{\xi}\right) V_5}{3V_2}=-\mathcal{K}^4_{1334,14}, \\
		\mathcal{K}^4_{2434,23}=\frac{V_2V_5 \sin^2\theta \left(t \ddot{\xi}+\dot{\xi}\right) \left(t \left(t \ddot{\xi}-4 \dot{\xi}\right)+6 \xi\right)}{3 t^2}=-\mathcal{K}^4_{2334,24}.	
\end{cases}
		\end{eqnarray}
		
		Based on the expressions provided in equations \eqref{RC}, \eqref{CR}, \eqref{Q(g,C)} and \eqref{Q(Ric,C)}, we derive the following results:



 \begin{pr}\label{pr2} The IBH spacetime with $6\xi-4t\dot{\xi}+t^2\ddot{\xi} \neq 0,$ $3\xi-t^2\dot{\xi}+2t\xi(t\xi-\dot{\xi}) \neq 0$ and $6\xi-2t^2\dot{\xi}+4t\xi(t\ddot{\xi}-\dot{\xi} \neq 0$ conforms to the prescribed curvature conditions :

 \begin{enumerate}[label=(\roman*)]
 				
%
%
	 
	  \item $\ R\cdot R-Q(S,R)=L_1Q(g,C),$ where $L_1=\frac{2(3\xi-2t\xi\dot{\xi}-t^2\dot{\xi}+2t^2\xi\ddot{\xi})}{tV_5},$
	  \item $C\cdot R-R\cdot C=L_2Q(g,R)+L_3Q(S,R),$ where $L_2=\frac{(2\dot{\xi}V_1+t\xi\ddot{\xi})V_5}{3t^2(3\xi-t^2\dot{\xi}+2t\xi(t\ddot{\xi-\dot{\xi}}))},
	  	L_3=\frac{V_1 V_5}{6\xi-2t^2\dot{\xi}+4t\xi(t\ddot{\xi}-\dot{\xi})},$
	   \item $C\cdot R-R\cdot C=L_4Q(g,C)+L_5Q(S,C),$ where $L_4=\frac{2(2\dot{\xi}+t\ddot{\xi})}{3t^2},L_5=1.$
	  \end{enumerate}	
\end{pr}


The projective curvature tensor $P_{pqrs}$ considering its symmetry properties has the following non-zero components

$$\begin{array}{c}
	P_{1212}=-\frac{2V_3}{3t^3}=-P_{1221},  \\
	P_{1313}=-\frac{V_3}{3V_2}=-\frac{1}{\sin^2\theta}P_{1414}, \\
    P_{1331}=-\frac{V_8}{3V_2}=-\frac{1}{\sin^2\theta}P_{1441},\\
	P_{2323}=\frac{V_2V_3}{3t^2}=\frac{1}{\sin^2\theta}P_{2424},\\
	P_{2332}=\frac{V_2V_8}{3t^2}=\frac{1}{\sin^2\theta}P_{2442}, \\
	P_{3434}=-\frac{2t\sin^2\theta V_8}{3}=-P_{3443}.
\end{array}$$

The concircular curvature tensor $W_{pqrs}$ considering its symmetry properties, has non-zero components and are presented as follows:

$$\begin{array}{c}
W_{1212}=-\frac{12\xi-14t\dot{\xi}+5t^2\ddot{\xi}}{6t^3},\\
	W_{1313}=-\frac{V_5}{6V_2}=\frac{1}{\sin^2\theta}W_{1414}, \\
	W_{2323}=\frac{V_2V_5}{6t^2}=\frac{1}{\sin^2\theta}W_{2424}. 
\end{array}$$

The conharmonic curvature tensor $K_{pqrs}$ considering its symmetry properties has the following non-zero components
	
$$\begin{array}{c}
	K_{1212}=-\frac{2V_1}{t^3}=-\frac{1}{t\sin^2\theta}K_{3434},\\ 
	K_{1313}=-\frac{V_9}{2V_2}=\frac{1}{\sin^2\theta}K_{1414}, \\
	K_{2323}=\frac{V_2V_9}{2t^2}=\frac{1}{\sin^2\theta}K_{2424}. \ \ 
\end{array}$$

In \cite{SDHK_interior_2020}, the following properties have been fulfilled by IBH spacetimes:
\begin{thm}[\cite{SDHK_interior_2020}] The IBH spacetime obeys the following curvature properties:
\begin{enumerate}[label=(\roman*)]
\item $R \cdot Z = N_1 Q(g, Z)$, \ $N_1 = \frac{\xi - t \dot{\xi}}{t^3}$,
\item $C \cdot Z = N_2 Q(g, Z)$, \ $N_2 = \frac{6 \xi - 4t \dot{\xi} + t^2 \ddot{\xi}}{6 t^3}$,
\item $W \cdot Z = N_2 Q(g, Z)$,
\item $K \cdot Z = N_3 Q(g, Z)$, \ $N_3 = \frac{2\xi + t^2 \ddot{\xi}}{2t^3}$,
\item $P \cdot S = N_1 Q(g, S)$,
\item $P \cdot Z = N_1 Q(g, Z) - \frac{1}{3}Q(S, Z)$,
where $Z$ is any one of $R, S, C, W, K$ and $P$.
\end{enumerate}
\end{thm}
According to the given theorem, the following results hold:  
\begin{itemize}
    \item[(a)] If \(\xi = C_1 t\), then the conditions \(R \cdot Z = 0\) and \(P \cdot S = 0\) are satisfied.  
    \item[(b)] If \(\xi = C_1 t^2 + C_2 t^3\), it follows that \(C \cdot Z = 0\) and \(W \cdot Z = 0\).  
    \item[(c)] When  
    \[
    \xi = \sqrt{t} \left( C_2 \cos\left(\frac{\sqrt{7}}{2} \log t\right) + C_1 \sin\left(\frac{\sqrt{7}}{2} \log t \right) \right),
    \]  
    the equation \(K \cdot Z = 0\) holds true.  
\end{itemize}
Here, \(C_1\) and \(C_2\) represent arbitrary constants.
 
Further, in \cite{SDHK_interior_2020} the following characteristics of IBH spacetimes has been revealed:
\begin{thm}[\cite{SDHK_interior_2020}] The IBH spacetime realizes the following curvature properties:
\begin{enumerate}[label=(\roman*)]
\item $C \cdot K = W \cdot K$, \ \ \ \ \ $C \cdot K = W \cdot C$, \ \ \ \ \ $C \cdot C = W \cdot C$,
\item $W \cdot K = W \cdot C$, \ \ \ \ \ $C \cdot K = C \cdot C$ \ ((ii) follows from (i)),
\item $W \cdot K = C \cdot C$ \ ((iii) follows from (i) and (ii)),
\item $C \cdot W = W \cdot R$, 
\item $R \cdot S = P \cdot S$, \ \ \ \ \ $C \cdot S = W \cdot S$,
\item $N_3 \ R \cdot K = N_1 \ K \cdot C$,
\item $R \cdot W - W \cdot R = N_5 \ Q(g, R)$, \ $N_5 = - \frac{2\dot{\xi} + t \ddot{\xi}}{6t^2} = \frac{r}{12}$,
\item $C \cdot K - K \cdot C = N_6 \ Q(g, C)$, \ $N_6 = - \frac{2 \dot{\xi} + t \ddot{\xi}}{3t^2}  = \frac{r}{6}$,
\item $C \cdot R - Q(S, C) = N_7 \ Q(g, C)$, \ $N_7 = \frac{3 \xi + t \dot{\xi} + 2t^2 \ddot{\xi}}{3t^3}$,
\item $R \cdot R - Q(S, R) = N_8 \ Q(g, C)$, \ $N_8 = \frac{6\xi^2 - 4 t \xi \dot{\xi} - 2t^2 \dot{\xi}^2 + 4t^2 \xi \ddot{\xi}}{t^3(6 \xi - 4 t \dot{\xi} + t^2 \ddot{\xi})}$,
\item $N_6 N_9 \ R \cdot W + N_1 N_{10} \ Q(S, W) = N_1 N_{11} \ Q(S, R)$, \ $N_9 = - 3t (6\dot{\xi}(\xi- t\dot{\xi})+t(t\dot{\xi}+3\xi)\ddot{\xi})$,
 \ \ \ \ \ $N_{10} = 6(-t^2 \dot{\xi}^2+2 t \xi(t \ddot{\xi}-\dot{\xi})+3 \xi^2)$, \ $N_{11} = t^2(-t^2 \ddot{\xi}^2+2 \dot{\xi}^2+2t\dot{\xi} \ddot{\xi})+6t\xi(t \ddot{\xi}-4 \dot{\xi})+18 \xi^2$,
\item $N_3 \ R \cdot K + N_{12} \ K \cdot R + N_3 \ Q(S, K)=0$, \ $N_{12} = \frac{-2\xi + 4 t \dot{\xi} + t^2 \ddot{\xi}}{2t^3}$,
\item $N_3 N_{13} \ R \cdot K - N^2_1 N_2 \ K \cdot R + N_1 N_2 N_3 \ Q(S, R) = 0$, \ $N_{13} = \frac{3\xi^2 - 2 t \xi \dot{\xi} - t^2 \dot{\xi}^2 + 2t^2 \xi \ddot{\xi}}{3 t^6}$,
\item $N_2 \ C \cdot W - N_7 \ W \cdot C = N_2 \ Q(S, C)$,
\item $N_2 N_{14} \ C \cdot W - \frac{1}{18t^6}N_{11} \ W \cdot C = N^2_2 \ Q(S, W)$, \ \ $N_{14} = \frac{3\xi - 5t \dot{\xi} - t^2 \ddot{\xi}}{3t^3}$,
\item $N_{15} \ C \cdot K + N^2_2 \ Q(S, K) = N_2 N_{16} \ Q(S, C)$, \ $N_{15} = \frac{4t^2\dot{\xi}+4t^3\dot{\xi}\ddot{\xi}+t^4 \ddot{\xi}^2)}{3t^6}$, \ \ \ $N_{16} = \frac{2\xi -4t \dot{\xi}-t^2 \ddot{\xi}}{2t^3}$,
\item $18t^6 N_3 N_{11} \ W \cdot K - N^2_2 N_{14} \ K \cdot W + N^2_2 N_3 \ Q(S, W) = 0$,
\item $N_1 N_3 \ W \cdot K + N_2 N_{12} \ K \cdot W + N_2 N_3 \ Q(S, K)=0$,
\item Roter type condition with $R = \frac{\phi}{2}S\wedge S + \mu g\wedge S + \eta G$, where 
$$
\phi = - \frac{ 6 t \xi - 4 t^2 \dot{\xi} + t^3 \ddot{\xi}}{(t \ddot{\xi} - 2 \dot{\xi})^2}, 
\mu = -\frac{6 \xi \dot{\xi} - 6 t \dot{\xi}^2 + 3 t \xi \ddot{\xi} + t^2 \xi \ddot{\xi}}{t(t \ddot{\xi} - 2 \dot{\xi})^2}, 
\eta = -\frac{2(4 \xi \dot{\xi}^2 - 4 t \dot{\xi}^3 + 2 t \xi \dot{\xi} \ddot{\xi} + t^2 \xi \ddot{\xi}^2)}{t^3(t \ddot{\xi} - 2 \dot{\xi})^2}.
$$
\end{enumerate}
\end{thm}
Based on the given theorem, the following conclusions can be drawn:  
\begin{itemize}
    \item[(a)] When \(\xi = C_1 t\), the relation \(R \cdot K = K \cdot C\) holds.  
    \item[(b)] If \(\xi = -\frac{C_1}{t} + C_2\), then the conditions \(R \cdot W = W \cdot R\) and \(C \cdot K = K \cdot C\) are satisfied.  
    \item[(c)] For  
    \[
    \xi = t^{\frac{1}{4}} \left( C_2 \cos\left(\frac{\sqrt{23}}{4} \log t\right) + C_1 \sin\left(\frac{\sqrt{23}}{4} \log t\right) \right),
    \]  
    it follows that \(C \cdot R = Q(S, C)\).  
\end{itemize}
Here, \(C_1\) and \(C_2\) denote arbitrary constants.

From \eqref{RR},\eqref{Q(g,R)},\eqref{Q(Ric,R)} \eqref{RC},\eqref{CR}, \eqref{Q(g,C)} and \eqref{Q(Ric,C)}, we can further state the following:
\begin{thm}
The IBH spacetime with $6\xi-4t\dot{\xi}+t^2\ddot{\xi} \neq 0,$ $3\xi-t^2\dot{\xi}+2t\xi(t\xi-\dot{\xi}) \neq 0$ and $6\xi-2t^2\dot{\xi}+4t\xi(t\ddot{\xi}-\dot{\xi} \neq 0$ reveals the following geometric structures:

	\begin{enumerate}[label=(\roman*)]

		\item the relation $R\cdot R$ is linearly dependent with $Q(S,R)$ and $Q(g,C),$ i.e.,$\ R\cdot R-Q(S,R)=L_1Q(g,C),$ where $L_1=\frac{2(3\xi-2t\xi\dot{\xi}-t^2\dot{\xi}+2t^2\xi\ddot{\xi})}{tV_5},$
		
		\item the commutator $C\cdot R-R\cdot C$  can be expressed as a linear combination of the tensors $Q(g,R)$ and $Q(S,R)$ as well as $Q(g,C)$ and $Q(S,C)$, i.e.,\\ $(a)$$C\cdot R-R\cdot C=L_2Q(g,R)+L_3Q(S,R),$ where $L_2=\frac{(2\dot{\xi}V_1+t\xi\ddot{\xi})V_5}{3t^2(3\xi-t^2\dot{\xi}+2t\xi(t\ddot{\xi-\dot{\xi}}))},
			  	L_3=\frac{V_1 V_5}{6\xi-2t^2\dot{\xi}+4t\xi(t\ddot{\xi}-\dot{\xi})},$\\
			    $(b)$$C\cdot R-R\cdot C=L_4Q(g,C)+L_5Q(S,C),$ where $L_4=\frac{2(2\dot{\xi}+t\ddot{\xi})}{3t^2},L_5=1.$
		
		\item the conformal $2$-forms are recurrent in IBH spacetime for the $1$-form $({\frac{2\dot{\xi}-2t\ddot{\xi}+t^2{\xi}^3}{6\xi-4t\dot{\xi}+t^2\ddot{\xi}},0,0,0})$, where $6\xi-4t\dot{\xi}+t^2\ddot{\xi} \neq 0,$
		 
%

	\item it is a $2$-quasi Einstein manifold for $\mathscr{X} = \frac{2\dot{\xi}}{t^2} $,

		\item it is an Ein $(2)$ manifold, as it fulfills $S^2+ \lambda_1S+\lambda_2 g=0, $  where $\lambda_1=\frac{2\dot{\xi}+t\ddot{\xi}}{t^2}$ and  $\lambda_2=\frac{2\dot{\xi}\ddot{\xi}}{t^3},$
     \item it is generalized quasi-Einstein, i.e., 
     $S=\alpha g+\beta \eta\otimes\eta+\epsilon(\eta\ast\delta+\delta\ast\eta),$ where $\alpha=-\frac{\ddot{\xi}}{t},\beta=1,\epsilon=1,\delta_1=0,\delta_2=0,\delta_3=-\frac{1+2\dot{\xi}-t\ddot{\xi}}{2},\delta_4=\frac{\iota\sin\theta(1+2\dot{\xi}-t\ddot{\xi})}{2}$	and $\eta_1=0,\eta_2=0,\eta_3=1,\eta_4=-\iota\sin\theta,$	
		
		
		 
		\item the universal form of tensors that are compatible with $R$, $C$, $P$, $W$ and $K$ can be derived as follows: 
		
		$$
		\left(
		\begin{array}{cccc}
			\mathscr{Y}_{11} &\mathscr{Y}_{12} & 0 & 0 \\
			\mathscr{Y}_{12} & \mathscr{Y}_{22} & 0 & 0 \\
			0 & 0 & \mathscr{Y}_{33} & \mathscr{Y}_{34} \\
			0 & 0 & \mathscr{Y}_{34} & \mathscr{Y}_{44}
		\end{array}
		\right),
		$$
		where $\mathscr{Y}_{ij}$ are arbitrary scalars,

	\end{enumerate}
	
\end{thm}

\begin{rem}
From analyzing the components of various curvatures, we can deduce that the IBH spacetime does not support the following geometric structures:
	\begin{enumerate}[label=(\roman*)]
		\item since $\nabla W\neq 0,$ it follows that $\nabla K\neq 0$, $\nabla C\neq 0$, $\nabla R\neq 0$ and $\nabla P\neq 0$,
		

%
%
%
		
		
		

		\item the curvature $2$-forms associated with $W,$ $P,$ $K$ and $R$ are not recurrent. 
		
		
	\end{enumerate}  
\end{rem}

\section{\bf Nature of the Energy momentum tensor of the Interior black hole spacetime}

Albert Einstein developed the physics of spacetime using a system of equations to characterize its geometric features in the general theory of relativity. The governing equation is given by
$$S-\frac{{\kappa}}{2}g+\Lambda g= \frac{8\pi G}{c^4}T,$$ 
where $S$ represents the Ricci curvature, ${\kappa}$ is the scalar curvature, $T$ denotes the energy-momentum tensor of the spacetime, $\Lambda$ signifies the cosmological constant, $G$ is the gravitational constant and $c$ is the speed of light in a vacuum. This formula provides a mathematical explanation of how energy and matter influence the curvature of spacetime, carrying significant consequences for our comprehension of gravity and the universe at large.



If we assume $\frac{8\pi G}{c^4}=1$, then the components of the energy-momentum tensor are expressed as:

$$\begin{array}{c}
	T_{11}=\frac{2\dot{\xi}}{tV_2}, \ \
	T_{22}=\frac{2V_2\dot{\xi}}{t^3}, \\
	T_{33}=t\ddot{\xi}=\frac{1}{\sin^2\theta}T_{44}, \\

\end{array}$$

and the trace of the energy momentum tensor is given by $-\frac{2(2\xi+t\ddot{\xi})}{t^2}.$\\
Let $\mathscr{J}^{1} = R \cdot T,$ $\mathscr{J}^{2} = C \cdot T,$ $\mathscr{J}^{3} = W \cdot T,$ $\mathscr{J}^{4} = K \cdot T$ and $\mathscr{J}^5 = Q(g, T).$ We now provide the non-zero components of $\mathscr{J}^{1},$ $\mathscr{J}^{2},$ $\mathscr{J}^3$ $\mathscr{J}^{4},$ and $\mathscr{J}^5,$ considering their symmetries, as follows:

\begin{eqnarray}\label{RT}
\begin{cases}
\mathscr{J}^1_{1313}=\frac{V_1V_4}{t^2V_2}=\frac{1}{\sin^2\theta}\mathscr{J}^1_{1414},\\\mathscr{J}^1_{2323}=-\frac{V_2V_1V_4}{t^4}=\frac{1}{\sin^2\theta}\mathscr{J}^1_{2424};\\			
	\end{cases}
			\end{eqnarray}

\begin{eqnarray}\label{CT}
\begin{cases}
\mathscr{J}^2_{1313}=\frac{V_4V_5\sin^2\theta}{6t^2V_2}=\mathscr{J}^2_{1414},\\ \mathscr{J}^2_{2323}=\frac{V_2V_4V_5}{6t^4}=\frac{1}{\sin^2\theta}\mathscr{J}^2_{2424};\\					
	\end{cases}
			\end{eqnarray}
			
\begin{eqnarray}\label{WT}
\begin{cases}
\mathscr{J}^3_{1313}=\frac{V_4V_5}{6t^2V_2}=\frac{1}{\sin^2\theta}\mathscr{J}^3_{1414},\\\mathscr{J}^3_{2323}=\frac{V_2V_4V_5}{6t^4}=\frac{1}{\sin^2\theta}\mathscr{J}^3_{2424};\\				
	
	\end{cases}
			\end{eqnarray}
			
\begin{eqnarray}\label{KT}
\begin{cases}
\mathscr{J}^4_{1313}=\frac{V_4V_9}{2t^2V_2}=\frac{1}{\sin^2\theta}\mathscr{J}^4_{1414},\\\mathscr{J}^4_{2323}=\frac{V_2V_4V_9}{2t^4}=\frac{1}{\sin^2\theta}\mathscr{J}^4_{2424};\\				
	
	\end{cases}
			\end{eqnarray}
					
\begin{eqnarray}\label{gT}
\begin{cases}
\mathscr{J}^5_{1313}=\frac{V_4}{V_2}=\frac{1}{\sin^2\theta}\mathscr{J}^5_{1414},\\\mathscr{J}^5_{2323}=\frac{V_2V_4}{t}=\frac{1}{\sin^2\theta}\mathscr{J}^5_{2424}.\\				
	
	\end{cases}
			\end{eqnarray}
\indent The following conclusions can be drawn from the components stipulated above.:
\begin{thm} The characteristics of the energy momentum tensor in the IBH spacetime are explored as:
\begin{enumerate}[label=(\roman*)]
\item  $R\cdot  T=\frac{\xi-t\dot{\xi}}{t^3} Q(g,T)$, 
		\item $C\cdot T=\frac{6\xi-t(4\dot{\xi}-t\ddot{\xi})}{6t^3} Q(g,T)$,
		
		\item $W\cdot  T=\frac{6\xi-t(4\dot{\xi}-t\ddot{\xi})}{6t^3} Q(g,T)$,
			\item $K\cdot T =\frac{2\xi+t^2\ddot{\xi}}{2t^3} Q(g,T)$.
	\end{enumerate}
\end{thm}
The above theorem leads to the the following remark.

\begin{rem}
The energy momentum tensor of the IBH spacetime obeys semisymmetric type conditions for the following cases: 
   \begin{enumerate} [label=(\roman*)]
    \item if $\xi=C_1t$ then $R\cdot T=0,$
 \item if $\xi=t^2(C_2 +C_3 t)$ then $C\cdot T=0$ and $W\cdot T=0,$
 \item if $\xi=C_4t^3+\frac{C_5}{t}$ then $K\cdot T=0,$
   \end{enumerate}
   where $C_i$'s are constants for $i=1,2,3,4,5$.
\end{rem}

%

%

\section{\bf Curvature inheritance and collineation admitted by Interior black hole spacetime}


The Lie subalgebra of $\chi(M)$ is the set of all Killing vector fields on a manifold $\mathcal{M}_{\nu},$ represented by $\mathcal{K}(M)$. It is known that $\mathcal{K}(M)$ can possess not more than $\frac{\nu(\nu+1)}{2}$ linearly independent Killing vector fields. The manifold $\mathcal{M}_{\nu}$ is referred to as a maximally symmetric space when $\mathcal{K}(M)$ contains precisely $\frac{\nu(\nu+1)}{2}$ linearly independent Killing vector fields. If a manifold $\mathcal{M}_{\nu}$  has constant scalar curvature, then it is maximally symmetric. The IBH spacetime is not maximally symmetric. In the IBH spacetime
 $\frac{\partial}{\partial z},$  and $\frac{\partial}{\partial \phi}$ are Killing vector fields. Also, note that, in IBH spacetime ${\frac{\partial}{\partial t}}$ and $\frac{\partial}{\partial \theta}$ are non-Killing vector fields.
Let $\mathcal G_{ij}=\pounds_ Xg_{ij}$, for $i,j=1,2,3,4$, and for the non-Killing vector field $X=(1-\frac{2\xi}{t})\frac{\partial}{\partial t}$. The non-vanishing components of $\mathcal G_{ij}$  can be computed as follows.
%
$$\begin{array}{c}
 \mathcal G_{11}=\frac{2V_1}{tV_2},\
\mathcal G_{22}=-\frac{2V_1V_2}{t^3},\\
	\mathcal G_{33}=2V_2=\frac{1}{\sin^2\theta} \mathcal G_{44}.
\end{array}$$
	\begin{pr}\label{pr3} The IBH spacetime with respect to the non-Killing vector field $X=(1-\frac{2\xi}{t})\frac{\partial}{\partial t},$ do not admit the following structures:
	\begin{enumerate}[label=(\roman*)]
	\item Ricci soliton, Yamabe soliton, 
	\item gradient Ricci soliton, generalized Ricci soliton,
	\item $\eta$-Ricci soliton, $\eta$-Yamabe soliton,
	\end{enumerate}
	\end{pr}
Let $\mathscr{B}^1=g\wedge g$, $\mathscr{B}^2=S\wedge S$, $\mathscr{B}^3=g\wedge S$, then the non-zero components of $\mathscr{B}^1, \mathscr{B}^2,$ and $\mathscr{B}^3$ are computed as follows:
 \begin{eqnarray}\label{gag}
  \begin{cases}
  \mathscr{B}^1_{1212}=\mathscr{B}^1_{1221}=\mathscr{B}^1_{2112}=\mathscr{B}^1_{2121}=2;\\
  \mathscr{B}^1_{1313}=-\mathscr{B}^1_{1331}=-\frac{1}{\sin^2\theta}\mathscr{B}^1_{1441}=\frac{1}{\sin^2\theta}\mathscr{B}^1_{1414}\\
  =\mathscr{B}^1_{3131}=-\mathscr{B}^1_{3113}=-\frac{1}{\sin^2\theta}\mathscr{B}^1_{4114}=\frac{1}{\sin^2\theta}\mathscr{B}^1_{4141}=-\frac{2t^3}{V_2};\\
  \mathscr{B}^1_{2323}=\mathscr{B}^1_{2332}=-\mathscr{B}^1_{3223}=-\mathscr{B}^1_{3232}\\
  =\frac{1}{\sin^2\theta}\mathscr{B}^1_{2424}=\frac{1}{\sin^2\theta}\mathscr{B}^1_{2442}=\frac{1}{\sin^2\theta}\mathscr{B}^1_{4224}=\frac{1}{\sin^2\theta}\mathscr{B}^1_{4242}=-2tV_2;\\
  \mathscr{B}^1_{3434}=-\mathscr{B}^1_{3443}=-\mathscr{B}^1_{4334}=\mathscr{B}^1_{4343}=2t^4\sin^2\theta;\\
  \end{cases}
 \end{eqnarray}
 \begin{eqnarray}\label{sas}
   \begin{cases}
   \mathscr{B}^2_{1212}=-\mathscr{B}^2_{1221}=-\mathscr{B}^2_{2112}=\mathscr{B}^2_{2121}=\frac{2\ddot{\xi}^2}{t^2};\\
   \mathscr{B}^2_{1313}=-\mathscr{B}^2_{1331}=-\frac{1}{\sin^2\theta}\mathscr{B}^2_{1441}=\frac{1}{\sin^2\theta}\mathscr{B}^2_{1414}\\
   =\mathscr{B}^2_{3131}=-\mathscr{B}^2_{3113}=-\frac{1}{\sin^2\theta}\mathscr{B}^2_{4114}=-\frac{1}{\sin^2\theta}\mathscr{B}^2_{4141}=-\frac{4\dot{\xi}\ddot{\xi}}{V_2};\\
   \mathscr{B}^2_{2323}=-\mathscr{B}^2_{2332}=-\mathscr{B}^2_{3223}=\mathscr{B}^2_{3232}\\
   =\frac{1}{\sin^2\theta}\mathscr{B}^2_{2424}=\frac{1}{\sin^2\theta}\mathscr{B}^2_{2442}=-\frac{1}{\sin^2\theta}\mathscr{B}^2_{4224}=\frac{1}{\sin^2\theta}\mathscr{B}^2_{4242}=\frac{4V_2\dot{\xi}\ddot{\xi}}{t^2};\\
   \mathscr{B}^2_{3434}=-\mathscr{B}^2_{3443}=-\mathscr{B}^2_{4334}=\mathscr{B}^2_{4343}=-8\dot{\xi}^2\sin^2\theta;\\
   \end{cases}
  \end{eqnarray} 
  \begin{eqnarray}\label{gas}
    \begin{cases}
    \mathscr{B}^3_{1212}=-\mathscr{B}^3_{1221}=-\mathscr{B}^3_{2112}=\mathscr{B}^3_{2121}=-\frac{2\ddot{\xi}}{t};\\
    \mathscr{B}^3_{1313}=-\mathscr{B}^3_{1331}=-\frac{1}{\sin^2\theta}\mathscr{B}^3_{1441}=\frac{1}{\sin^2\theta}\mathscr{B}^3_{1414}\\
    =\mathscr{B}^3_{3131}=-\mathscr{B}^3_{3113}=-\frac{1}{\sin^2\theta}\mathscr{B}^3_{4114}=\frac{1}{\sin^2\theta}\mathscr{B}^3_{4141}=\frac{tV_{11}}{t};\\
    \mathscr{B}^3_{2323}=-\mathscr{B}^3_{2332}=-\mathscr{B}^3_{3223}=\mathscr{B}^3_{3232}\\
    =\frac{1}{\sin^2\theta}\mathscr{B}^3_{2424}=-\frac{1}{\sin^2\theta}\mathscr{B}^3_{2442}=-\frac{1}{\sin^2\theta}\mathscr{B}^3_{4224}=\frac{1}{\sin^2\theta}\mathscr{B}^3_{4242}=-\frac{V_2V_{11}}{t};\\
    \mathscr{B}^3_{3434}=-\mathscr{B}^3_{3443}=\mathscr{B}^3_{4334}=-\mathscr{B}^3_{4343}=4t^2\dot{\xi}\sin^2\theta.\\
    \end{cases}
   \end{eqnarray}  
   Again for the non-Killing vector field $X=(1-\frac{2\xi}{t})\frac{\partial}{\partial t},$ we consider $\pounds_{X}R=\mathcal{N}^1$, $\pounds_{X}S=\mathcal{N}^2$, $\pounds_{X}C=\mathcal{N}^3$, $\pounds_{X}W=\mathcal{N}^4$, $\pounds_{X}K=\mathcal{N}^5$ and $\pounds_{X}P=\mathcal{N}^6,$ where $\pounds_{X}$ denotes the Lie derivative with respect to the vector field $X.$ Then non-vanishing components of $\mathcal{N}^1,$ $\mathcal{N}^2,$ $\mathcal{N}^3,$ $\mathcal{N}^4,$ $\mathcal{N}^5,$ and $\mathcal{N}^6$ are computed as follows:
   \begin{eqnarray}\label{LXR}
        \begin{cases}
        \mathcal{N}^1_{1212}=-\mathcal{N}^1_{1221}=-\mathcal{N}^1_{2112}=\mathcal{N}^1_{2121}=\frac{1}{t^5}((-10+t(3+4\dot{\xi}))(2\xi+V_4)-t^3V_2\xi^{(3)});\\
        \mathcal{N}^1_{1313}=-\mathcal{N}^1_{1331}=-\mathcal{N}^1_{3113}=\mathcal{N}^1_{3131}=\frac{1}{t^2V_2}(V_1(V_{1}+2t\dot{\xi})+\ddot{\xi});\\
        \mathcal{N}^1_{1414}=-\mathcal{N}^1_{1441}=-\mathcal{N}^1_{4114}=\mathcal{N}^1_{4141}=\frac{\sin^2\theta}{t^2V_2}(\xi V_{10}(1+6\dot{\xi}-2t\ddot{\xi})-t^2(\dot{xi(1+V_{11})}));\\
        \mathcal{N}^1_{2323}=-\mathcal{N}^1_{2332}=\frac{1}{\sin^2\theta}\mathcal{N}^1_{2424}=-\frac{1}{\sin^2\theta}\mathcal{N}^1_{2442}=\frac{V_2}{t^2}(4\xi^2+t^2(\dot{\xi}-V_4)-t\xi(1+6\xi+2t\ddot{\xi}));\\
        \mathcal{N}^1_{3223}=-\mathcal{N}^1_{3232}=\frac{1}{\sin^2\theta}\mathcal{N}^1_{4224}=-\frac{1}{\sin^2\theta}\mathcal{N}^1_{4242}=\frac{V_2}{t^4}(-4\xi^2-t^2\dot{\xi}(1+V_{11})-t\xi(1+6\xi-2t\ddot{\xi}));\\
        \mathcal{N}^1_{3434}=-\mathcal{N}^1_{3443}=-\mathcal{N}^1_{4334}=\mathcal{N}^1_{4343}=\frac{2\sin^2\theta}{t}(V_2(\xi+t\dot{\xi}));\\
        \end{cases}
       \end{eqnarray}
       \begin{eqnarray}\label{LXS}
            \begin{cases}
            \mathcal{N}^2_{11}=\frac{V_{10}+2t\dot{\xi}\ddot{\xi}-tV_2\xi^{(3)}}{t^2V_2};\\
            \mathcal{N}^2_{22}=-\frac{V_2}{t^4}(V_{10}-2t\dot{\xi}\ddot{\xi}-tV_2\xi^{(3)});\\
            \mathcal{N}^2_{33}=-\frac{2V_2\ddot{\xi}}{t}=\frac{1}{\sin^2\theta}\mathcal{N}^2_{44};\\
            \end{cases}
           \end{eqnarray} 
   \begin{eqnarray}\label{LXC}
        \begin{cases}
        \mathcal{N}^3_{1212}=-\mathcal{N}^3_{1221}=-\mathcal{N}^3_{2112}=\mathcal{N}^3_{2121}=\frac{1}{3t^5}(-60\xi^2-2t\xi(9+34\dot{\xi}-7t\ddot{\xi}+t^2\xi^{(3)})+t^2(-5t\ddot{\xi}\\+2\dot{\xi}(7+8\dot{\xi}-2t\ddot{\xi})+t^2\ddot{\xi}));\\
        \mathcal{N}^3_{1313}=\mathcal{N}^3_{1331}=-\mathcal{N}^3_{3113}=\mathcal{N}^3_{3131}\\
        =-\frac{1}{\sin^2\theta}\mathcal{N}^3_{4114}=\frac{1}{\sin^2\theta}\mathcal{N}^3_{4141}=\frac{1}{\sin^2\theta}\mathcal{N}^3_{1441}=-\frac{1}{\sin^2\theta}\mathcal{N}^3_{1414}=-\frac{1}{6t^2V_2}(24\xi^2-2t\xi(3+16\dot{\xi}\\-4t\ddot{\xi}+t^2\xi^{(3)})+t^2(-3t\ddot{\xi}+2\dot{\xi}(3+4\dot{\xi}-t\ddot{\xi}+t^2\xi^{(3)})));\\
        \mathcal{N}^3_{2323}=-\mathcal{N}^3_{2332}=\frac{1}{\sin^2\theta}\mathcal{N}^3_{2424}=\frac{1}{\sin^2\theta}\mathcal{N}^3_{2442}\\
        =-\mathcal{N}^3_{3223}=\mathcal{N}^3_{3232}=-\frac{1}{\sin^2\theta}\mathcal{N}^3_{4224}=\frac{1}{\sin^2\theta}\mathcal{N}^3_{4242}=\frac{V_2}{6t^4}(24\xi^2-2t\xi(3+16\dot{\xi}\\-4t\ddot{\xi}+t^2\xi^{(3)})+t^2(-3t\ddot{\xi}+2\dot{\xi}(3+4\dot{\xi}-t\ddot{\xi}+t^2\xi^{(3)})));\\
        \mathcal{N}^3_{3434}=\mathcal{N}^3_{3443}=-\mathcal{N}^3_{4334}=\mathcal{N}^3_{4343}=-\frac{V_2\sin^2\theta}{3t}(6\xi+t(-2\dot{\xi}+t(-\ddot{\xi}+t\xi^{(3)})));\\
        \end{cases}
       \end{eqnarray} 
   \begin{eqnarray}\label{LXW}
        \begin{cases}
        \mathcal{N}^4_{1212}=-\mathcal{N}^4_{1221}=-\mathcal{N}^4_{2112}=\mathcal{N}^4_{2121}=\frac{1}{3t^5}(-120\xi^2-2t\xi(18+92\dot{\xi}-29t\ddot{\xi}+5t^2\xi^{(3)})+t^2(-19t\ddot{\xi}\\+4\dot{\xi}(10+14\dot{\xi}-5t\ddot{\xi})+5t^2\xi^{(3)}));\\
                \mathcal{N}^4_{1313}=\mathcal{N}^4_{1331}=-\mathcal{N}^4_{3113}=\mathcal{N}^4_{3131}\\
                =-\frac{1}{\sin^2\theta}\mathcal{N}^4_{4114}=\frac{1}{\sin^2\theta}\mathcal{N}^4_{4141}=\frac{1}{\sin^2\theta}\mathcal{N}^4_{1441}=-\frac{1}{\sin^2\theta}\mathcal{N}^4_{1414}=-\frac{1}{6t^2V_2}(24\xi^2-2t\xi(3+16\dot{\xi}\\-4t\ddot{\xi}+t^2\xi^{(3)})+t^2(-3t\ddot{\xi}+2\dot{\xi}(3+4\dot{\xi}-t\ddot{\xi}+t^2\xi^{(3)})));\\
                \mathcal{N}^4_{2323}=-\mathcal{N}^4_{2332}=\frac{1}{\sin^2\theta}\mathcal{N}^4_{2424}=\frac{1}{\sin^2\theta}\mathcal{N}^4_{2442}\\
                =-\mathcal{N}^4_{3223}=\mathcal{N}^4_{3232}=-\frac{1}{\sin^2\theta}\mathcal{N}^4_{4224}=\frac{1}{\sin^2\theta}\mathcal{N}^4_{4242}=\frac{V_2}{6t^4}(24\xi^2-2t\xi(3+16\dot{\xi}\\-4t\ddot{\xi}+t^2\xi^{(3)})+t^2(-3t\ddot{\xi}+2\dot{\xi}(3+4\dot{\xi}-t\ddot{\xi}+t^2\xi^{(3)})));\\
                \mathcal{N}^4_{3434}=\mathcal{N}^4_{3443}=-\mathcal{N}^4_{4334}=\mathcal{N}^4_{4343}=-\frac{V_2\sin^2\theta}{3t}(-12\xi+t(-8\dot{\xi}+t(-5\ddot{\xi}+t\xi^{(3)})));\\
        \end{cases}
       \end{eqnarray} 
    \begin{eqnarray}\label{LXK}
         \begin{cases}
         \mathcal{N}^5_{1212}=\mathcal{N}^5_{2121}=\frac{2}{t^5}(4V_1^2-V_2(3\xi+t(V_4-\dot{\xi})));\\
         \mathcal{N}^5_{1221}=\mathcal{N}^5_{2112}=\frac{1}{t^5}(20\xi^2+2t^2(\dot{\xi}(3+4\dot{\xi})-t\ddot{\xi})+2t\xi(-3-14\dot{\xi}+2t\ddot{\xi}));\\
         \mathcal{N}^5_{1313}=-\mathcal{N}^5_{1331}=-\frac{1}{\sin^2\theta}\mathcal{N}^5_{1441}=\frac{1}{\sin^2\theta}\mathcal{N}^5_{1414}\\
         =\mathcal{N}^5_{3131}=-\mathcal{N}^5_{3113}=-\frac{1}{\sin^2\theta}\mathcal{N}^5_{4114}=\frac{1}{\sin^2\theta}\mathcal{N}^5_{4141}=-\frac{1}{2t^2V_2}(8\xi^2+t^2(\dot{\xi}(2-2t\ddot{\xi})+t(\ddot{\xi+t\xi^{(3)}})\\-2\xi(t+4t\xi+t^3\xi^{}{(3)}));\\
         \mathcal{N}^5_{2323}=-\mathcal{N}^5_{2332}=\mathcal{N}^5_{3223}=-\mathcal{N}^5_{3232}\\
         =\frac{1}{\sin^2\theta}\mathcal{N}^5_{2424}=-\frac{1}{\sin^2\theta}\mathcal{N}^5_{2442}=\frac{1}{\sin^2\theta}\mathcal{N}^5_{4224}=-\frac{1}{\sin^2\theta}\mathcal{N}^5_{4242}=-\frac{V_2}{2t^4}(8\xi^2+t^2(\dot{\xi}(2-2t\ddot{\xi})+t(\ddot{\xi+t\xi^{(3)}})\\-2\xi(t+4t\xi+t^3\xi^{}{(3)}));\\
         \mathcal{N}^5_{3434}=-\mathcal{N}^5_{3443}=-\mathcal{N}^5_{4334}=\mathcal{N}^5_{4343}=-\frac{2\sin^2\theta}{6t}(-12\xi+t(-8\dot{\xi}+t(5\ddot{\xi}+t\xi^{(3)})));\\
         \end{cases}
        \end{eqnarray} 
  \begin{eqnarray}\label{LXP}
       \begin{cases}
       \mathcal{N}^6_{1212}=-\mathcal{N}^6_{1221}=\mathcal{N}^6_{2112}=-\mathcal{N}^6_{2121}=\frac{1}{3t^5}(-60\xi^2-2t^2(-4t\ddot{\xi}+\dot{\xi}(9-4V_4+4\dot{\xi})+t^2\xi^{(3)}\\+2t\xi(9+42\dot{\xi}+2t(-6\ddot{\xi}+t\xi^{(3)}))));\\
       \mathcal{N}^6_{1313}=-\frac{1}{\sin^2\theta}\mathcal{N}^6_{1441}=\frac{1}{\sin^2\theta}\mathcal{N}^6_{1414}=\frac{1}{3t^2V_2}(-12\xi^2-t^2(-t\ddot{\xi}+\dot{xi}(3+6\dot{xi}-2t\ddot{\xi})+t^2\xi^{(3)})\\-2t\xi(9+42\dot{\xi}+2t(t\xi^{(3)}-6\ddot{\xi})));\\
       \mathcal{N}^6_{1331}=\mathcal{N}^6_{3131}=\frac{1}{3t^2V_2}(V_1(-12\xi+t(2\dot{\xi}+3)-t^2V_2\ddot{\xi}));\\
       \mathcal{N}^6_{3113}=\mathcal{N}^6_{4114}=\frac{\sin^2\theta}{3t^2V_2}(12\xi^2+t^2(-2t\ddot{\xi}+\dot{\xi}(3+6\dot{\xi}-2t\ddot{\xi})+t^2\xi^{(3)}-t\xi(3+18\xi-6t\ddot{\xi}+2t^2\xi^{(3)})));\\
       \mathcal{N}^6_{4141}=\frac{\sin^2\theta}{3t^2V_2}(-12\xi^2+t^2(-t\dot{\xi}+2\dot{\xi})+t\ddot{\xi})+t\xi(3+14\dot{\xi}2t\ddot{\xi})));\\
       \mathcal{N}^6_{2323}=-\mathcal{N}^6_{2332}=-\mathcal{N}^6_{3223}=\mathcal{N}^6_{3232}\\
       =\frac{1}{\sin^2\theta}\mathcal{N}^6_{2424}=-\frac{1}{\sin^2\theta}\mathcal{N}^6_{2442}=-\frac{1}{\sin^2\theta}\mathcal{N}^6_{4224}=\frac{1}{\sin^2\theta}\mathcal{N}^6_{4242}=\frac{V_2}{2t^4}(8\xi^2+t^2(\dot{\xi}(2-2t\ddot{\xi})+t(\ddot{\xi+t\xi^{(3)}})\\-2\xi(t+4t\xi+t^3\xi^{}{(3)}));\\
       \mathcal{N}^6_{3434}=-\mathcal{N}^6_{3443}=-\mathcal{N}^6_{4334}=\mathcal{N}^6_{4343}=-\frac{2\sin^2\theta V_2}{3t}(-3\xi+t(-\dot{\xi}+t(-\dot{\xi}+\ddot{\xi}))).\\
       \end{cases}
      \end{eqnarray} 
\indent From the components exhibited in equations \eqref{gag}, \eqref{sas}, \eqref{gas},\eqref{LXR},\eqref{LXS},\eqref{LXC}, \eqref{LXW}, \eqref{LXK} and \eqref{LXP}, we obtain the following results:                        
\begin{thm}
For the non-Killing vector field $X=(1-\frac{2\xi}{t})\frac{\partial}{\partial t},$ the IBH spacetimes explores the following properties:
\begin{enumerate}[label=(\roman*)]
\item $\pounds_{X}R=R+\mu_1 g\wedge g+\mu_2g\wedge S+\mu_3S\wedge S,$ i.e., generalized curvature inheritance,\\
where,
\begin{align*}
 \mu_1=\frac{4\dot{\xi}^2V_1+t\xi\ddot{\xi}V_{11}}{t^3V_4^2}+\frac{1}{t^5V_4^2}(V_1(12t\dot{\xi}^2-40\xi\dot{\xi}^2+16t\dot{\xi}^3)+V_2(2t\xi\dot{\xi}\ddot{\xi}+4t^2\dot{\xi}^2\ddot{\xi}-t^2\xi\ddot{\xi}^2+t^3\xi\ddot{\xi}^2-2t^3\dot{\xi}^2\xi^{(3)})\\-4t\dot{\xi}\ddot{\xi}(\xi^2-t^2\dot{\xi})),
 \end{align*}
$$\mu_2=\frac{6\dot{\xi}V_1+t\ddot{\xi}V_{12}}{tV_4^2}+\frac{1}{t^3V_4^2}(V_1(14t\dot{\xi}-48\xi\dot{\xi}20t\dot{\xi}^2-t^2\ddot{\xi}-6t^2\dot{\xi}\ddot{\xi})+V_2(4t^2\dot{\xi}\ddot{\xi}+t^3\ddot{\xi}^2-2t^3\dot{\xi}\xi^{(3)}))$$
$$\mu_3=\frac{tV_5}{2v_4^2}+\frac{1}{2tV_4^2}(V_{10}(6\xi-10t\dot{\xi})+4t\dot{\xi}(t-3t\dot{\xi})+t^2\ddot{\xi}(5t-14\xi+4t\dot{\xi})-V_2t^3\xi^{(3)})$$
\item $\pounds_{X}S=\mu_4S+\mu_5 g,$ i.e., generalized Ricci inheritance,\\
where $\mu_4=\frac{1}{t^3V_4}(3t\ddot{\xi}-8\xi\ddot{\xi}+2t\dot{\xi}\ddot{\xi}-t\xi^{(3)}V_2)$\\
$\mu_5=\frac{1}{t^3V_4}(2V_4V_2+2\dot{\xi}(3t\ddot{\xi}-8\xi\ddot{\xi}+2t\dot{\xi}\ddot{\xi}-t\xi^{(3)}V_2))$
\item $\pounds_{X}C=\mu_6C+\mu_7 g\wedge g+\mu_8g\wedge S,$ i.e., special case of generalized conformal curvature inheritance,
where $$\mu_6=\frac{1}{t^2V_5}(-6V_1V_{10}-8t\dot{\xi}V_1+3t^2\ddot{\xi}(t+3\xi)-t^2\ddot{\xi}(t-2t\dot{\xi})t^3\xi^{(3)}V_2)$$
$$\mu_7=\frac{1}{t^5V_4}(-12t\xi\dot{\xi}(1-t\ddot{\xi})+(8t^2\dot{\xi}^2+2t^3\dot{\xi}\ddot{\xi})(1+\dot{\xi})+36\xi\dot{\xi}(\xi-\dot{\xi})-6t\xi\ddot{\xi}(t-3\xi)-t^3\ddot{\xi}^2(t+3\xi-t\dot{\xi}))$$
$$\mu_8=\frac{-2}{3t^3V_4}(18\xi V_1+(1+\dot{\xi})(t^3\ddot{\xi}-4t^2\dot{\xi})-3t\xi(t\ddot{\xi}-2))$$
\item $\pounds_{X}W=\mu_9W+\mu_{10} g\wedge g+\mu_{11}g\wedge S,$ i.e., special case of generalized concicular curvature inheritance,
where $$\mu_9=\frac{1}{t^2V_5}(V_{10}(10t\dot{\xi}-6\xi)-4t\dot{\xi}(\xi-3t\dot{\xi})-t^2\ddot{\xi}(5t+14\xi-4t\dot{\xi}))$$
\begin{align*}
\mu_{10}=\frac{1}{48t^7V_4V_5\sin^2\theta}(V_9-14V_1)((6\xi-10t\dot{\xi}) V_{10}+4t\dot{\xi}(\xi-3t\dot{\xi})+t^2\ddot{\xi}(5t-14\xi+4t\dot{\xi})-t^3\xi^{(3)}V_2)\\(\frac{2\sin^2\theta\dot{\xi}}{3t^3}(-120\xi^2+t^2(19t\ddot{\xi}-4\dot{\xi}(10+14\dot{\xi}-56\ddot{\xi}-5t^2\xi^{(3)}+2t\xi(18+92\dot{\xi}-29t\ddot{\xi}+5t^2\xi^{(3)}))\\-\frac{1}{3t^2}(V_2\ddot{\xi}\sin^2\theta(-12\xi+t(-8\dot{\xi}+t(5\ddot{\xi}+t^{(3)})))))))
\end{align*}
\begin{align*}
\mu_{11}=\frac{1}{2t^2V_5}((6\xi V_{10}-\dot{\xi}(10t^2+44\xi-12t^2\dot{\xi})+t^2\ddot{\xi}(5t-14\xi+4t\dot{\xi})-t^2\xi{(3)}V_2)-\frac{1}{6t^3V_4}(24\xi(t-3\xi)\\-16t\dot{\xi}(t-5\xi)+4t\dot{\xi}(\xi-7t\dot{\xi})+t^2\ddot{\xi}(7t-24\xi+10t\dot{\xi})-3t^3\xi^{(3)}V_2))
\end{align*}
\item $\pounds_{X}K=\mu_{12}K+\mu_{13} g\wedge g+\mu_{14}g\wedge S,$ i.e., special case of generalized conharmonic curvature inheritance,
where $$\mu_{12}=\frac{-1}{t^2V_5}(V_1(6V_{10}+8t\dot{\xi})+t^2\ddot{\xi}(3t-8\xi+2t\dot{\xi})t^2\xi^{(3)}V_2)$$
\begin{align*}
\mu_{13}=-\frac{-1}{t^5V_4}((6t\dot{\xi}-20\xi\dot{\xi})V_1+(t\xi\ddot{\xi}+t^2\xi\ddot{\xi}-t^3\ddot{\xi})V_2+8t\dot{\xi}(\xi-t\dot{\xi}^2))-\frac{V_1}{t^5V_5}(6\xi V_{10}-6\dot{\xi}(t^2-4\dot{\xi})+8\dot{\xi}(\xi-t^2\dot{\xi}) \\ +t^2\ddot{\xi}(3t-8\xi+2t\dot{\xi})-t^3\xi^{(3)}V_2)
\end{align*}
$$\mu_{14}=\frac{4}{t^3V_4}(-\xi(t-3\xi)+t\dot{\xi}(V_{10}+t\dot{\xi}))$$
\item $\pounds_{X}P=\mu_{15}P+\mu_{16} g\wedge g+\mu_{17}g\wedge S+\mu_{18}S\wedge S,$ i.e., generalized Weyl projective curvature inheritance,\\
where $$\mu_{15}=\frac{-1}{t^2V_4}(4\dot{\xi}V_1-t\ddot{\xi}V_{10}-2t^2\xi\ddot{\xi}+t^2\xi^{(3)}V_1)$$
\begin{align*}
\mu_{16}=\frac{1}{3t^5V_4^3}((-12\dot{\xi}^2(\xi-\dot{\xi})+2tv\ddot{\xi}V_8-t^2\ddot{\xi}V_{12})(4\dot{\xi}V_1+t\ddot{\xi}(V_{10}+2t\dot{\xi})-t^2\xi^{(3)}V_2)-\frac{1}{3t^5V_4^2}(-36t\dot{\xi}^2(\xi-\dot{\xi})\\+(120\xi\dot{\xi}-48t\dot{\xi}^3)V_1-6t\xi\dot{\xi}\ddot{\xi}V_{10}-10t^2\dot{\xi}^2\ddot{\xi}(t-\xi)+2t^2\dot{\xi}^2\ddot{\xi}(\xi-2t\ddot{\xi})+(3t^2\xi\ddot{\xi}-3t^2\dot{\xi}\ddot{\xi}^2-3t^3\dot{\xi}\ddot{\xi}^2-t^4\ddot{\xi}^3\\+4t^3\dot{\xi}^2\xi^{(3)}+2t^4\dot{\xi}\ddot{\xi}\xi^{(3)})V_2)+4t^3\ddot{\xi}^2(\xi\dot{\xi}-t))
\end{align*}
\begin{align*}
\mu_{17}=\frac{-1}{3t^3V_4^3}(18\dot{\xi}V_1+t\ddot{\xi}(t^2\ddot{\xi}+t\dot{\xi}+9\xi)(4\dot{\xi}V_1+t\ddot{\xi}(V_{10}+2t\dot{\xi})-t^2\xi{(3)}V_2)-\frac{1}{3t^3V_4^2}(-42t\dot{\xi}+144\xi\dot{\xi}-60t\dot{\xi}^2\\+3t^2\ddot{\xi}+14t^2\dot{\xi}\ddot{\xi}+2t^3\ddot{\xi}^2)V_1+(t^4\ddot{\xi}\xi^{(3)}+6t^3\dot{\xi}\xi^{(3)}-4t^3\ddot{\xi}-12t^2\dot{\xi}\ddot{\xi})V_2)
\end{align*}
$$\mu_{18}=\frac{{-1}}{3tV_4^3}((18\xi\dot{\xi}-26t\dot{\xi}^2+20t^2\dot{\xi}\ddot{\xi}-5t^3\ddot{\xi}^2-9t^2\xi\xi^{(3)}+3t^2\dot{\xi}\xi^{(3)})V_2+(8t\dot{\xi}^2-8t^2\dot{\xi}\ddot{\xi}+2t^3\ddot{\xi}^2)V_1)$$
where $\xi$ is a function of $t$.
\end{enumerate}
\end{thm}
%
%
%
In particular, if $\xi$ is constant (i.e., $\dot{\xi}=0=\ddot{\xi}$),
then  for $X=\frac{\partial}{\partial t}$ and $\frac{\partial}{\partial \theta},$ the Lie derivative $\pounds_XS=0,$ i.e., it reveals Ricci collineation with respect to $\frac{\partial}{\partial t}$, $\frac{\partial}{\partial \theta},$
and if $\xi=\xi(t),$  a function of $t,$ then  for $X=\frac{\partial}{\partial t}$ and $\frac{\partial}{\partial \theta},$ $\pounds_XS\neq 0$ and hence it does not exhibit Ricci collineation.

\section{\bf Interior black hole spacetime Vs Kiselev black hole spacetime  }
%
In $2003$, Kiselev introduced a well-known black hole metric for toy models \cite{Kiselev1}. This metric is a static, spherically symmetric solution to the Einstein field equations. It is a straightforward, singularity-free black hole spacetime in general relativity. Kiselev model has been cited over $200$ times, both directly and indirectly, with more than $150$ of the referring works being published.
In spherical coordinates $(t, r, \theta, \phi)$, the KBH metric \cite{ESKiselev2024} is given by:
\[
ds^2 = -\left(1-\frac{2m}{r}-\frac{\sigma}{r^{1+3\omega}}\right)dt^2 + \left(1-\frac{2m}{r}-\frac{\sigma}{r^{1+3\omega}}\right)^{-1}dr^2 + r^2\left(d\theta^2 + \sin^2\theta d\phi^2\right),
\]
where $m$ represents the mass of the black hole and
$\sigma$ is the quintessence's parameter.  The Schwarzchild and Reissner-Nordström-de Sitter-like black hole solutions were derived by Kiselev, and they lie in the quintessence phase in the range of $-1\leq \omega \leq -\frac{1}{3},$ where the parameter $\omega$ characterizes the state of the matter surrounding the black hole.

 A comparison of the curvature properties between IBH spacetime and KBH spacetime is outlined as follows:\\

\noindent\textbf{Similarities:}
\begin{enumerate}[label=(\roman*)]
	
	\item both spacetimes admit pseudosymmetry, as a consequence conformally, concircularly, conharmonically and projectively pseudosymmetric,
   \item  both spacetimes are of $Ein(2)$ manifolds,
	
	\item conformal $2$-forms are recurrent for both spacetimes,
	
	\item in both spacetimes, the  coordinate  vector field $\frac{\partial}{\partial \phi}$ is Killing,
	
	\item the vector field $\frac{\partial}{\partial \theta}$ is non-Killing in both spacetimes,
	
	
   \item for both spacetimes the energy momentum tensor is pseudosymmetric due to conformal, concircular and conharmonic curvature tensor,
   \item both spacetimes are belonging to the Roter type.
\end{enumerate}
Nonetheless, they distinct in the following characteristics:\\

\noindent\textbf{Dissimilarities:}

\begin{enumerate}[label=(\roman*)]
	\item the projective $2$-forms of IBH spacetime are not recurrent while KBH spacetime shows such recurrency,
	
	\item IBH spacetime is a $2$-quasi Einstein manifold whereas
	KBH spacetime is $3$-quasi Einstein,
	
	
		
	 
	
	\item in IBH spacetime the coordinate vector fields $\frac{\partial}{\partial t}$ is non-Killing whereas in KBH spacetime it is Killing,
	
	
	\item Ricci $1$-forms are not recurrent in IBH but it admits for KBH spacetime.
	
\end{enumerate}
\section{\bf  Acknowledgment}



The second author greatly acknowledges to The University Grants Commission, Government of India for the award of Junior Research Fellow. All the algebraic computations of Section $3-5$ are performed by a program in Wolfram Mathematica developed by the first
author A. A. Shaikh.

\section{\bf Declarations}

{\bf Data availability:} Data sharing not applicable to this article as no data sets were used/generated or analyzed during the current study.\\

{\bf Conflict of interests:} The authors have declared no conflicts of interest that are relevant to the content of this article.\\

{\bf Funding:} No funding was received to assist with the preparation of this manuscript.\\


%


\end{document}